**On the Electromagnetic Design of a $\beta_g$= 0.61, 650 MHz Superconducting Radiofrequency Cavity**

Arup Ratan Jana and Vinit Kumar

Materials and Advanced Accelerator Science Division

Raja Ramanna Center for Advanced Technology, Indore, India

E-mail: arjana@rrcat.gov.in, arup.jana@gmail.com

**Abstract:** We present the calculations for electromagnetic design of a $\beta_g$ = 0.61, multi-cell superconducting radiofrequency cavity for the Indian Spallation Neutron Source (ISNS) project. Geometry of the mid-cells is optimized using a step by step, one-dimensional optimization technique. This is followed by the optimization of end-cells, which is done to achieve the required field flatness, as well as to avoid trapping of higher order modes. Calculations of the threshold beam current for the excitation of regenerative beam break up instability excited by the dipole modes is also presented, which is followed by wake field calculations and estimation of its effects. Specific aspects relevant to design calculations for medium $\beta_g$ cavities, compared to high $\beta_g$ cavities are highlighted in the paper. Finally, studies are performed for the static as well as dynamic Lorentz Force Detuning (LFD), based on which the stiffness design of the cavity is optimized.

## I. Introduction

Multi-cell elliptic superconducting radiofrequency(SRF) cavities are used for efficient acceleration of high power charged particle beam for a wide range of velocities, typically corresponding to $\beta = 0.5$ to 1, where $\beta$ is the particle speed in unit of speed of light [1-4]. There is a plan to build an Indian Spallation Neutron Source(ISNS)[2] for neutron based multidisciplinary research, for which SRF cavities will be used to accelerate a beam of $H^-$ particles from ~ 160 MeV to 1 GeV. For optimum performance, the multi-cell SRF cavities are designed for $TM_{010} - \pi$ mode of operation, where the cell length is $\beta_g\lambda/2$, where $\lambda$ is the free space wavelength of radio-frequency (RF) wave used for acceleration, and $\beta_g$ is the value of $\beta$ for which perfect synchronism exists between the RF wave and the charge particle for maximum acceleration. There will however be a range of $\beta$, for which a cavity corresponding to a particular value of $\beta_g$ can be used. For the ISNS project, two families of such cavities will be used – the first family corresponding to $\beta_g = 0.9$ that will be used to accelerate the $H^-$ beam from 500 MeV to 1 GeV, and the second family corresponding to $\beta_g = 0.61$ that will accelerate the $H^-$ beam from 160 MeV to 500 MeV. Calculations have been earlier performed for the electromagnetic design of a 5-cell, 650 MHz, $\beta_g = 0.9$ elliptic SRF cavity [2]. In this paper, we perform calculations for the electromagnetic design of a 5-cell, 650 MHz, $\beta_g = 0.61$ elliptic SRF cavity.

We start with the geometry optimization studies for $\beta_g = 0.61$ elliptic SRF cavity. In a multi-cell elliptic cavity, in general, the end-cells have different geometry compared to intermediate cells, which are called the mid-cells. We first optimize the geometry of mid-cells, for which we follow a step by step, one dimensional optimization technique developed earlier for the $\beta_g = 0.9$ cavities [2]. This is followed by optimization of the geometry of end-cells. Two criteria are used for this – achieving maximum field flatness, and ensuring that there is no trapped higher order mode (HOM) [2-4]with significant strength. The second requirement leads to asymmetric end-cells for $\beta_g$ = 0.61 cavity, which is a new feature compared to the earlier designed $\beta_g$ = 0.9 cavity [2]. After having completed the geometry optimization of the mid-cells as well as end-cells, a justification is given for the choice of iris radius, which is in terms of the maximum acceptable value of cell to cell coupling coefficient needed to ensure the desired criterion on field flatness in the cavity.

Next, the higher order dipole modes supported by the cavity have been studied and their strengths in terms of $R/Q$, where $R$ is the effective shunt impedance and $Q$ is the quality factor is calculated. If the beam current is higher than a threshold value, the dipole mode generated by the beam can grow exponentially due to beam cavity interaction giving rise to regenerative beam break up (BBU) instability[4-6]. We calculate the threshold current for this instability for the prominent dipole modes supported by the optimized geometry. Unlike the calculations performed for the $\beta_g$ = 0.9 cavities in Ref. 2, here more elaborate calculations are performed for a range of values of $\beta$, for which a detailed discussions on effect of particle velocity and phase velocity is presented. We then perform the calculations of the wakefield[3-4,7] generated by charged particle bunches in the $\beta_g$ = 0.61 cavity, and estimate its effect on the beam dynamics, as well as on the cavity heat dissipation. Unlike the calculations performed for the $\beta_g$ = 0.9 cavities in Ref. 2, here the beam is non-ultra-relativistic, and therefore more elaborate calculations are needed, which are performed for longitudinal wakefield.

Finally, Lorentz Force Detuning (LFD)[2-4,8] generated by the radiation pressure due to RF power inside the cavity is evaluated for the optimized design. Since ISNS will be a pulsed machine, the radiation pressure will be pulsed and therefore the study on dynamic LFD[8-9] is important. The static LFD however gives a good idea about dynamic behavior also. We have therefore first performed the calculations of static LFD, and optimized the thickness of the cavity wall, the helium vessel design and range for the location of stiffener ring in order to suppress the LFD. The fine tuning of the location of the stiffener ring is done on the basis of dynamic LFD.

This paper is organized as follows. In Section II, we present the electromagnetic design studies for the optimization of the cavity geometry. A discussion on the general features related to cavity geometry as well as the requirements based on superconducting RF properties is first presented. Detailed calculations to obtain the optimized mid cell geometry are described in the Section II-A. Section II-B discusses the optimization of the end cell geometry to get the desired field flatness. Discussion on the calculations to choose the optimum iris radius to achieve desired field flatness with maximum shunt impedance is presented in Section II-C. Next, in the Section III, we present the calculation of $R/Q$ for monopole HOMs and optimize the end cell geometry to ensure that there is no trapped mode with significant value of $R/Q$. This is followed by the calculation of $R/Q$ for dipole HOMs for the optimized cavity, and also the threshold current for regenerative BBU instability. Section IV discusses the calculations of wakefield and its effect on the beam dynamics, as well as calculation of cryogenic load generated due to wakefield. LFD studies and its implication on proper choice of the cavity stiffening strategy are discussed in Section V. Finally, some discussions and conclusions are presented in Section VI.

## II. ELECTROMAGNETIC DESIGN OPTIMIZATION OF THE CAVITY GEOMETRY

We start the design study with optimization of the cavity geometry. For the cavity geometry, we have followed the typical TESLA-type cavity shape [3], which has evolved over several years to minimize the multipacting problems[2-3,9-10]. Figure 1 shows the schematic of half-cell of this type of cavity, which is constructed by joining two elliptic arcs with a straight line. The 3-D cavity shape is a figure of revolution around the beam axis, obtained using the contour shown in this figure. In a multi-cell cavity, in general, the geometry of the end cells is different compared to the intermediate cells, which are called the mid-cells. We first optimize the geometry of mid cells. As seen in Fig. 1, the geometry of a half cell is described by seven independent parameters - the half-cell length $L$, iris ellipse radii $a$ and $b$, equator ellipse radii $A$ and $B$, iris radius $R_{iris}$ and equator radius $R_{eq}$. The wall angle $\alpha$ can be calculated as a derived parameter from these seven independent parameters. For the mid-cells, we choose $L = \beta_g\lambda/2$, and $R_{eq}$ is tuned to achieve the resonant frequency as 650 MHz[2]. We perform the optimization of the remaining parameters by using the step by step procedure discussed in Ref. 2.

Before we start with the calculation of optimized geometry parameters of the mid cells, we discuss the criteria and constraints. Our target is to achieve the maximum possible acceleration field $E_{acc}$ in the cavity. Here the two constraints *i.e.* the peak surface magnetic field $B_{pk}$ that leads to the breakdown in the superconducting properties and the peak surface electric field $E_{pk}$ that leads to field emission, limit the maximum achievable value of $E_{acc}$[9]. Therefore, the aim of the optimization study is to minimize the value of $B_{pk}/E_{acc}$, keeping a satisfactory value for $E_{pk}/E_{acc}$. The peak magnetic field $B_{pk}$, developed on the inner surface of the cavity wall must be less than the critical magnetic field and for a niobium SRF cavity operating at 650 MHz, 70 mT can be taken as the value for the $B_{pk}$ as a safe margin[2,11]. Similarly, from the field emission point of view, the maximum tolerable value of peak electric field $E_{pk}$ on the inner surface of the cavity wall must be ≤ 40MV/m[2,12]. For our design, the target value of $E_{pk}/E_{acc} \leq 2.355$ and $B_{pk}/E_{acc} \leq$ 4.56 mT(MV/m) will ensure an accelerating gradient ~ 15.5 MV/m.

*A. Optimization of the Mid Cell Geometry*

As discussed earlier, we first fix the half-cell length $L$ equal to $\beta_g\lambda/2$ and in our case, this is equal to 70.336 mm. We choose $R_{iris}$ = 44 mm, and give its justification later in Section II-C. Actually, the cell to cell coupling $k_c$ is decided mainly by $R_{iris}$. For a multi-cell cavity, this is very important and affects the field flatness. In addition, the constraint on $R_{iris}$ arises from the beam dynamics requirements. The wall angle $\alpha$ shown in Fig. 1 is another constraint for the mid cell geometry optimization and we keep it fixed at $88^0$ for the design presented here. In fact, by increasing the wall angle, it is possible to decrease the value of $E_{pk}/E_{acc}$. However, the upper limit on $\alpha$ comes from the manufacturing point of view and ease of chemical cleaning and processing.

Anotherfactor that affects the choice of $\alpha$is the LFD.We have observed that the value of LFD increases with $\alpha$.

For the mid cell geometry, the ratio $a/b$ is optimized to obtain the minimum value of $E_{pk}/E_{acc}$ in order to achieve the maximum $E_{acc}$, whereas the parameter $A/B$ is mainly decided by mechanical requirement like stiffness, rigidity *etc*. With this background, we start the optimization of the mid cell geometry. Amongst the seven independent variables, we have already fixed $R_{iris}$, $L$ and $\alpha$. Keeping $R_{eq}$ for tuning the resonant frequency of the cavity, we are left with only three independent variable in the cavity geometry — $a/b$, $A/B$ and $B$. We start our optimization with the simplest possible geometry, where $a/b = A/B = 1$ and in Fig. 2, we study the variation of $E_{pk}/E_{acc}$ and $B_{pk}/E_{acc}$ as a function of $B$. We start with $B \sim L/2$ and gradually increase its value and observe that up to a certain value of $B$, $E_{pk}/E_{acc}$ remains nearly constant and then increases rapidly afterwards. As seen in Fig. 2, this happens at $B$ ~ 55 mm, where the value of $E_{pk}/E_{acc}$ and $B_{pk}/E_{acc}$ become ~ 2.9 MV/m and ~ 4.312 mT/(MV/m) respectively. Note that the value of $E_{pk}/E_{acc}$ and $B_{pk}/E_{acc}$ are approaching our target values. The parameter $B$ affects the electromagnetic property of the cavity the least. Therefore, for our further optimization, we keep the value of $B$ fixed at 55.55 mm. Henceforth, we are only left with two independent variables, *i.e.*, $a/b$ and $A$.

In the second stage of our optimization, we calculated $E_{pk}/E_{acc}$ as a function of $A$, for a range of values of $a/b$ between 0.2 and1. For each value of $a/b$, simulation shows a monotonically increasing nature of $E_{pk}/E_{acc}$ as a function of $A$, and a monotonically decreasing nature of $B_{pk}/E_{acc}$ as a function of $A$. From Fig. 3(a), for each value of $a/b$, we obtain a particular value of $A$, which gives us $E_{pk}/E_{acc}$= 2.355, which is the target value. Using Fig. 3(b), we obtain $B_{pk}/E_{acc}$ for each of this pair of $a/b$ and $A$. We have thus corresponding to each set of $a/b$ and $A$, one geometry which gives $E_{pk}/E_{acc}$= 2.355 and a particular value of $B_{pk}/E_{acc}$. Amongst all these possible geometries, we now look for the geometry that gives the minimum value of $B_{pk}/E_{acc}$. Fig. 4(a) and Fig. 4(b) show the plot of $B_{pk}/E_{acc}$ as a function of $A$ and $a/b$ respectively for these selected geometries, From these plots, we obtain that $B_{pk}/E_{acc}$ is minimum for $A$ = 52.64 mm, and $a/b$ = 0.53 respectively, which corresponds to $a$ = 15.28 mm and $b$ = 28.83 mm for $\alpha = 88^0$. In Table-I, we show all the optimized geometry parameters for the mid cell that we have obtained. Table-II summarizes the corresponding RF parameters for the $TM_{010}$ like $\pi$ mode. We have obtained these values using the 2D code electromagnetic design code SUPERFISH[13].

*B. Preliminary Optimization for the End cell Geometry*

In this sub-section we discuss the calculations for optimization of the end-cell geometry. The end-cells see different boundary condition compared to the mid cells, because these are connected to the beam pipe at one end. In particular, the end half-cell is connected to the mid-cell at one end and to the beam pipe on the other end. This results in a slightly different resonant frequency of the multi-cell cavity (with end-cells and beam pipe) compared to the resonant frequency of mid-cells, if the end half-cell geometry is taken exactly the same as mid cell geometry. The end half-cell therefore needs to be tuned such that the cavity with end-cells resonate at the same frequency as that of mid-cells and the field flatness is maximized. For tuning the end half-cell geometry, we vary the half-cell length $L_e$, keeping all other parameters fixed. We find that for $L_e$ = 71.5495 mm, the resonant frequency is restored and the field flatness is maximized. We emphasize that this is only a preliminary optimization of the end half-cell geometry, since this needs to be further optimized to ensure that there are no trapped higher order modes with significant strength in the cavity, which will be done in Section III.

*C. Selection of the Iris Radius*

Optimum choice of the iris radius of a multi-cell cavity is governed by several considerations. Lower value of iris radius gives higher shunt impedance, which minimize the heat loss, whereas higher value of iris radius is preferred from beam dynamics considerations. Another very important implication of the choice of iris radius is that it affects the sensitivity of field flatness to geometrical errors in the cavity. The field flatness $\eta$ is defined as [14]

$$\eta = \left(1 - \frac{\sigma_E}{\mu}\right), \qquad (1)$$

where, $\sigma_E$ and $\mu$ are the standard deviation and mean value respectively of the maximum amplitudes of electric field values in different cells. It is desired that the value of $\eta$ for a multi-cell cavity should be close to unity to ensure ideal synchronization between the charged particle and the electromagnetic wave. Now, for an $N$-cell cavity, we can correlate $\eta$ with the inter cell coupling coefficient $k_c$ according to the following approximate relation [15],

$$(1 - \eta) \propto \frac{\sigma_f}{f} \frac{1}{k_c} N^{3/2}, \qquad (2)$$

where $\sigma_f/f$ is the relative *rms* error in the resonant frequency of different cells. It is seen in the above equation that for a given value of $N$ and the minimum achievable value of $\sigma_f/f$ (decided by achievable tolerance on cavity dimensions), the field flatness is decided by $k_c$, which is mainly dependent on the iris radius. For a multi-cell cavity, $k_c$ is obtained in terms of the resonant frequency for the $\pi$ and 0 modes denoted by $f_\pi$ and $f_0$ respectively, and using the following expression [10],

$$k_c = 2 \times \frac{(f_\pi - f_0)}{(f_\pi + f_0)}. \qquad (3)$$

For the 9-cell TESLA cavity, $k_c$ = 1.87 %, and this gives an acceptable value of field flatness [9]. Now, with an assumption that we obtain similar value for $\sigma_f/f$ as obtained in TESLA cavity by following similar machining and cavity processing procedure, we will obtain similar field flatness for a 5-cell cavity, if we choose $k_c$ = 0.775 %. Justification for the choice of $N$ = 5 will be presented in later section. From this argument, it is clear that in our case, we should choose iris radius such that $k_c \geq$ 0.77 %. Figure 5 shows a plot of $k_c$ as a function of $R_{iris}$. As expected, $k_c$ increases as $R_{iris}$ increases. The same plot also shows that $R/Q$ decreased as $R_{iris}$ increases. We should thus choose $R_{iris}$ such that $k_c$ approaches the desired value since increasing $R_{iris}$ beyond this value will be detrimental for $R/Q$. We notice that $R_{iris}$ should be more than 43 mm to ensure $k_c \geq$ 0.775 %. We thus choose $R_{iris}$= 44.0 mm.

**III. Higher Order Mode Studies**

In the previous section, we have presented an optimized geometry for $\beta_g$=0.61, 650 MHz, 5-cell SRF elliptical cavity. Although the geometry of the mid cell is finalized, the end cell geometry will be further fine-tuned after HOM studies in this section. Studies of dipole HOMs and the threshold current for regenerative beam break up instability will also be discussed in this section.

For a cylindrically symmetric pill box like structure, the resonating electromagnetic modes can be categorized into two classes — a) transverse electric or TE like mode, having no electric field component parallel to the beam axis except at the iris location or near the beam pipe attached to the cavity, and b) the transverse magnetic or TM like mode, having no magnetic field component parallel to axis except at the iris or near the beam pipe attached to the cavity [2,4,9]. As discussed in earlier sections, the cavity operates in $TM_{010}$- $\pi$ mode. In addition to the fundamental $TM_{010}$mode, the cavity supports higher order modes with higher resonant frequencies. These modes are generated due to interaction of the beam with the cavity, and limit the cavity performance. Amongst all these modes, mainly the monopole and dipoles can influence a well collimated on-axis beam significantly. Also, for these cavities, for the on-axis particle, the deflecting impulse due to the electric field in a TE mode exactly cancels the impulse due to the magnetic field [2,4]. Therefore, ideally, the TE modes do not produce significant contribution towards the HOMs. Hence, we restrict our HOM analysis to monopole and dipole modes only. We look at the HOMs having resonant frequency up to the upper cut off frequency of 2.0 GHz, which is decided by the beam pipe diameter. Relative strength of different HOMs present in a cavity will depend on their quality factor $Q$ and the shunt impedance $R$. In addition to the beam instability, HOMs produce heat on the cavity surface, which adds to the heat load to the helium bath and can lead to the thermal breakdown of the superconductivity.HOM studies therefore form an integral part of the cavity design.

For the fundamental accelerating mode, the energy gain $\delta W = qE_0Tl\,\mathrm{Cos}\,\theta$, which is the famous Panofsky formula[4]. Here, $q$ and $E_0$ stand for the charge of the particle and the average longitudinal electric field, $l$ is the cell length and $\theta$ is the phase of the particle. The term $T$ is the transit time factor, which is basically a measure of the synchronism between the phase velocity of the electromagnetic mode, and the particle velocity, and thus has strong dependence on $\beta$. The cavity shunt impedance $R$ has square dependence on the energy gain, and thus has strong dependence on $\beta$. We have calculated $R/Q$ for different monopole modes, as a function of $\beta$ using an electromagnetic code SLANS[16]. Figure 6 shows $R/Q$ for the five members of the first pass band, as a function of $\beta$. We notice that only for the $\pi$ mode at 649.99 MHz *i.e.* the operating mode frequency and $4\pi/5$ mode at 649.44 MHz, the $R/Q$ values are significant. Also, $R/Q$ for the $\pi$ mode dominates and shows a maximum of ~ 354 Ω at $\beta$ ~ 0.65. However, the$4\pi/5$mode has higher value of $R/Q$compared to the π mode, near the two end points of the $\beta$range in Fig. 6. This restricts the use of $\beta_g$ = 0.61 cavities to $\beta$ ranging from ~0.51 to ~0.76. Simulation shows that all the five modes of the second pass band have their $R/Q$ values significantly smaller compared to the $\pi$ mode or the $4\pi/5$ mode of the first pass band. We have then looked at the coupling coefficient $k_c$ for these pass bands, which is ~0.81% and ~1.65 % for the first and second pass bands respectively, and is smallest for the third pass band. Very small value of $k_c$ indicates the possibility of trapped modes. Figure 7 shows the plot the $R/Q$ value for the five monopole modes of the third pass band, as a function of $\beta$. We notice that except the mode at 1653.20 MHz, all the modes have their $R/Q$ values less than 10 Ω and, for the mode at 1653.2 MHz, $R/Q$ gradually increases with $\beta$ and approaches ~ 20 Ω at $\beta$ = 0.76. In order to check if there is mode trapping, we plot the axial electric field amplitude along the cavity length in Fig. 8(a).The trapped nature of the mode is clearly seen in this figure. The field amplitude is maximumnear the center of the cavity and the amplitude is relatively smaller at the cavity ends. The HOM couplers that out-couple the HOM from the cavity are generally put on the beam pipe only. As a consequence, it would not be possible to couple this mode out of the cavity. One way to solve this problem is to fine-tune the geometry of the end cell such that end cell individually resonates at 1653. 2 MHz for this particular HOM. In that case, the end cell will act as a resonator for this particular HOM, and the location of maximum field amplitude will shift to the end cell. We achieve this by changing the semi major axis *A* of that end cell from 52.64 mm to 52.25 mm, and at the same time, end cell length is so tuned that the fundamental mode frequency is fixed at 650.0 MHz. Fulfilling thes two criteria simultaneously, we finalise the end half-cell length at 71.24 mm. Figure 8(b) shows the plot of the axial electric field amplitude along the cavity length after fine-tuning of the end cell geometry. It is clearly seen that the mode is no longer trapped inside the cavity since the field maximum shifts from the center of the cavity towards the end cell. The HOM can now be out-coupled with an appropriate design of HOM coupler.Figure 9(a) and 9(b) are the field contours for the end-cell geometry for the accelerating mode *i.e.* the mode resonating at 650 MHz and for the mode resonating at 1653.2 MHz. Table-III presents the geometry parameters for the two asymmetric end cells. The end cell, modified in order to remove the trapped mode is called 'End Cell-A', whereas for the other end cell, called ‘End Cell-B’, we keep the same geometry as we have mentioned in the section-II. Table-IV summarizes the corresponding RF parameters for the $TM_{010}$-π mode for the 5-cell cavity with these end cell configurations.

For the final cavity geometry, we plot $T$ as a function of $\beta/\beta_g$ in Fig. 10 for different value of number of cells $N$ in the cavity. As discussed earlier, the operating range of $\beta$ in our case in is 0.51 to 0.76. As seen in this figure, if we choose $N = 5$, the transit time factor will always be more than 0.7 in the operating range of $\beta$. This justifies the choice of $N = 5$ in our design.

Next, we discuus the dipole modes supported by the optimized design of the cavity that we have presented. In a cylindrically symmetric pillbox like structure, the on-axis axial component of electric field is zero for a TM dipole mode, but it has a no-zero transverse gradient. The beam gets a transverse kick due to the field of a dipole mode, and experiences the off-axis axial electric field, due to which it will exchange energy with the dipole mode and under suitable circumstances, an instability, known as regenerative beam break up instability may build up if the beam current is higher than a threshold current $I_{th}$. It is important to ensure that operating beam current is less than $I_{th}$. It is therefore necessary to estimate the threshold current corresponding to prominent dipole modes. Strength of a dipole mode is given in terms of $R_{\perp}/Q$, which has the following expression[4]

$$\frac{R_{\perp}}{Q} = \frac{1}{\omega_n U_n}\left|\int_{z_s}^{z_e}\left(\frac{\partial E_z}{\partial r}\right)e^{i\frac{\omega_n z}{\beta c}}dz\right|^2. \tag{4}$$

The above expression is same as $k_n^2$ times the expression described in Ref [4] and is in agreement with the commonly followed convention. Here, $k_n$ , $\omega_n$ and $U_n$denote the wave vector, the angular frequency and the stored energy respectively of the $n^{th}$ dipole mode. Here $\partial E_z/\partial r$ is the transverse gradient of the axial electric field. For our optimized design, we calculated the $R_{\perp}/Q$ for the dipole modes up to the frequency slightly above the cut off frequency of the beam pipe, which is 2 GHz. As in the case of monopole modes, the dipole mode strength$R_{\perp}/Q$ also shows a strong dependency on $\beta$, and we observed that within the range of $\beta$ for which these cavities will be used, the most prominent dipoles modes are the member of the first dipole pass band. Particularly, the mode resonating at a frequency 961.982 MHz shows the maximum value of $R_{\perp}/Q$, which is approximately $4.43 \times 10^4\,\Omega/\mathrm{m}^2$. Other prominent dipole mode frequencies and their corresponding $R_{\perp}/Q$ and $Q$ values are given in Table V. Figure 11 shows the plot of $R_{\perp}/Q$ for the five modes of the first dipole pass band as a function of $\beta$.

Using the $R_{\perp}/Q$ data for these prominent dipole modes, we now estimate the threshold current $I_{th}$, which is given by the following expression [5-6]

$$I_{th} = \frac{\pi^3 (cp) k_n}{2q \times g(\psi) \times (R_{\perp}/L_{cav}) \cdot L_{cav}^2}. \tag{5}$$

Here, $p$ is the momentum of the charged particle and $L_{cav}$ ~ 0.710 m is the length of the cavity, $q$ denotes the magnitude of the charge of the particle being accelerated. In the above equation, the function $g(\psi)$ is a measure of the synchronization between the particle velocity $\beta c$ and the phase velocity $v_p$ of the particular mode. The functional form of $g(\alpha)$ is given by[17]

$$g(\psi) = \frac{1}{2}\left(\frac{\pi}{\psi}\right)^3 \times \left(1 - \cos\psi - \frac{\alpha}{2}\sin\psi\right). \quad (6)$$

Here, $\psi$ is the slip parameter and can be written as

$$\psi = \omega L_{cav}\left(\frac{1}{v_p} - \frac{1}{\beta c}\right). \quad (7)$$

In order to calculate $\psi$, we must estimate $v_p$ for which we plot the dispersion diagram of the first few dipole-pass bands in Fig. 12. In this plot, we obtained the resonant frequency values directly from the SLANS simulation results and to calculate the corresponding phase advance per cell $\varphi$ ,we follow the procedure given in the Ref 18 and use the following formula,

$$\cos\varphi(z) = \frac{E_z(r,z+L)+E_z(r,z-L)}{2E_z(r,z)}. \quad (8)$$

Since for a dipole mode, $E_z$ is zero on the axis, we evaluated the above expression at a slightly off axis position ($r_o$ ~1mm). Since the operating range of $\beta$ is between 0.51 and 0.76, in Fig. 12, only the modes lying in between the lines $v = 0.76c$ and $v = 0.51c$ may have exchange of energy with the beam. Hence we confined our threshold current calculation for these selected modes only. From the analysis, we observed that $I_{th}$ will be minimum for the dipole mode with a frequency of 965.852 MHz and the minimum value of $I_{th}$ is ~ 0.7 mA, which is sufficiently higher than that of the CW average beam current of 0.4 mA in our case. In Table V, we have presented the threshold currents corresponding to the prominent HOMs. From the above analysis, it clear that $I_{th}$ will be inversely proportional to the strength of the corresponding $R_\perp/Q$ for a particular mode. We however notice that for the first mode, for which $R_\perp/Q$ is maximum, $I_{th}$ is not the minimum. This is because for this mode, $v_p$ is ~ 0.76$c$ and the response function $g(\psi)$ for $\beta$ in the range 0.51 to 0.76 is significantly small and effectively increases the value of $I_{th}$ up to ~ 1.0 mA for this mode. The dependence of $g(\psi)$ on $\beta$ for the mode 965.852 MHz is shown in Figure 13. We have plotted the $I_{th}$ as a function of $\beta$ in the Fig. 14.

## IV. Wake field analysis

In the previous section, we discussed the monopole and dipole HOMs supported by the cavity and its implications on the cavity design. In this section, we will discuss about the wakefield generated by the interaction of the beam with the cavity and its implication on the performance of the cavity. There are two main implications of the wakefield. First, the wakefield produced by a particle is experienced by the trailing particles and they get accelerated or de-accelerated due to this, depending on their phases. This produces an energy spread. Second, the wakefield produces heat on the cavity walls and thus adds to the heat load.

First, we discuss the loss factor $\kappa$, which describes the energy loss of a unit point charge to a particular HOM supported by the cavity. This energy loss appears as the energy of the HOM in the cavity, and is function of the $R/Q$ for the mode, and has a strong dependence on $\beta$.The standard software, *e.g.*, ABCI[19] calculates the wakes and the loss factor for the ultra-relativistic cases only, assuming $\beta \sim 1$. In our case, the beam is not ultra-relativistic, and we have therefore calculated $\kappa$ for the entire range of $\beta$[20] by evaluating $R/Q$ and using the following formula[4]

$$\kappa_n = \frac{\omega}{4}\frac{R_n}{Q}, \tag{9}$$

where $\kappa_n$ is the loss factor for the $n^{\text{th}}$ mode. The $\beta$ dependence of the loss factor is due to $\beta$ dependence of $r_n/Q$ in the above formula. The above formulationis developed for a point charge and to calculate the same for a distribution of charges, the formula is modified to the following form[4]

$$\kappa_n = \left(\frac{\omega}{4}\frac{R_n}{Q}\right) \times e^{-\frac{1}{2}\left(\frac{\omega_n \sigma}{\beta c}\right)^2}. \tag{10}$$

Here $\sigma$ is the *rms* length of the Gaussian beam bunch, which is assumed to be around 5 mm in our calculations. The integrated loss factor $\kappa_{||}$ is obtained by summing the $\kappa_n$ value for all the values of $n$. The values of the integrated loss factor for $\beta \sim 1$ case, calculated using the above formula matches well with that obtained from ABCI,as seen in Fig. 15. The plot of the integrated loss factor with $\beta$ is shown in Fig. 16. Using Fig. 15, we get the integrated loss factor for $\beta$ = 0.61 as ~0.531 V/pC. Using this data we have calculated the parasitic heat loss on the cavity walls due to the excitation of wakefield by the beam. The power $P_0$ appearing as parasitic heat loss is related to $\kappa_{||}$ by the following equation [4]

$$P_0 = (\mathrm{PRR}) \times q^2 \times \kappa_{||}, \tag{11}$$

where PRR is the pulse repetition rate. In our case, the pulse length is 2 ms, macropulse current is 4 mA, and within a macropulse, the micropulses repeat at a frequency of 325 MHz. Thus charge $q$ per micropulse is ~ 12.3 pC.  Hence, for a PRR of 325 MHz, $q$ = 12.31 pC and $\kappa_{||}$ = 0.531 V/pC, we get power dissipation within the macropulse as 26.18 mW and the CW average will be 2.62 mW. Also we have estimated the limiting value for energy spread $\delta\xi$ induced in a single bunch due to wakefield, which is given by [4]

$$\delta\xi = -e \times q \times W_z^{/} \times s. \tag{12}$$

Here, $W_z^{/}$ is the slope of the longitudinal wake potential, which is estimated as ~ 0.209 V/pC/mm from Fig. 17, and $s$ is the bunch length, which is approximately 24 mm. Using the above equation, we obtain $\delta\xi$= 54.26 keV. The energy gain for a single cavity is ~11 MeV. Therefore, the relative spread in the beam energy will be ~ 0.49 %. The reported value for the beam energy spread in the TESLA cavity for a Gaussian beam bunch of ~ 1 mm length is 0.2 %. As described in Ref. 7, by proper adjustment of input beam phase, it is possible to minimize the energy spread induced by the wakefield.  We want to emphasize that we have calculate the relative spread in the beam energy, considering $\beta \rightarrow 1$. Hence, it is obvious that for $\beta$ less than unity, as in our case,we may have overestimated the energy spread.For the parasitic heat loss, we would like to point out that the above calculation is under the assumption that in a macropulse, wakefields from different micropulses are adding incoherently. This could be a valid approximation since the frequency of significant HOMs supported in the cavity is not an integral multiple of PRR.

## V. Lorenz Force Detuning Studies

We now discuss the effect of Lorentz pressure on the optimized design of the cavity, and estimate the detuning produced due to this, and discuss its remedy. Lorentz pressure is essentially the radiation pressure, originating from the electromagnetic energy fed into the cavity, which causes deformation in the cavity shape and perturbation in the cavity volume due to which the resonant frequency of the cavity changes. This shift in frequency is known as the Lorentz Force Detuning (LFD). Because of their extremely high $Q$ factor, the frequency bandwidth becomes small, and therefore this shift in frequency is not tolerable for SRF cavities. Because of the small bandwidth, even a small detuning may result in a large reflection of the input power from the cavity. For our design, the loaded $Q$ is expected to be around $5 \times 10^6$, which results in a bandwidth of ~ 130 Hz. Study of LFD therefore becomes important[2].

LFD is evaluated by analyzing the mechanical deformation in the cavity due to the Lorenz pressure given by the following expression[2,21]

$$P = \frac{1}{4}(\varepsilon_0 E^2 - \mu_0 H^2). \tag{13}$$

The electromagnetic field in the above expression has oscillations at RF frequency as well as at lower frequency corresponding to PRR. Mechanical modes of vibration do not respond to RF frequency. Hence, for a CW machine, the cavity effectively experiences a constant pressure, and for a pulsed machine, it experiences the pressure in pulses at the operating rep. rate of the machine. In the first case, a static deformation is produced in the cavity, which is known as static LFD. In the second case, the cavity deformation is oscillating in nature, and is known as dynamic LFD[8,22]. In this case, resonance may occur between the Lorentz pressure and the mechanical mode of the cavity, which needs to be carefully avoided.

Since we are designing the cavity for a pulsed operation, the dynamic LFD is going to be important. However, to start with, static LFD gives a good idea about the dynamic LFD. Based on the static analysis, we present some modifications in the geometry of the cavity accessories. We then discuss about the Fourier series representation of the pressure pulse, followed by the calculation of the structural modes of the cavity, using which we analyze the dynamic LFD[8].

In order to reduce this detuning, along with the proper selection of the wall thickness of the cavity, additional mechanical constraint in the form of stiffener ring [2,8] is introduced in between the consecutive cells of the cavity, as well as between the end-cell and the wall of the helium vessel. On the other hand, the cavity tuning that we attain by changing the cavity length poses a limitation on the stiffness of the cavity. Therefore, in order to finalize the design from static LFD considerations, both these criteria must be satisfied.

We have simulated a simple geometry of the vessel and performed the simulations described above for different values of stiffness of the helium vessel. A helium vessel of 504 mm diameter and having the end closure in "tori-flat" shape with a torus radius of 35 mm has been modeled for the study of static LFD, which is shown in Fig. 18[2]. The wall thickness of the helium vessel is taken as 4 mm and the material is taken as titanium. The cavity wall thickness is also taken as 4 mm. For these calculations, subroutines have been written in ANSYS[23] parametric design language and these calculations are presented for operation at acceleration gradient of ~ 15.5 MV/m. The reduction of the detuning strongly depends on the stiffness of the helium vessel also. We have therefore studied the LFD as a function of the stiffness of helium vessel, as well as the location of the stiffener ring. In order to compensate for the LFD, we have also studied the required range for tuning the resonant frequency by elongating the cavity. Figure 19 shows the LFD as a function of the stiffness of the helium vessel, where we observe that the LFD decreases significantly when we increase the stiffness from 1.9 kN/mm to 5 kN/mm and its absolute value decreases from ~ 4.5 kHz to ~ 2.1 kHz. Actually by increasing the thickness of the helium vessel and by changing the shape of the end closure from "tori-flat" to "tori-spherical", it is possible to attain this type of stiffness for the helium vessel. Next, we have studied the effect of the radial position of stiffeners on the static LFD for the same helium vessel stiffness, and this dependence is plotted in Fig. 20.

We finally study the possibility of compensating the LFD using a tuner. We apply a compensation by elongating the cavity, for which helium vessel needs to be displaced by 8.3 μm. Figure 21 shows the compensated static LFD as a function of the radial location of the stiffener ring. We notice in this figure that the tuning is possible by placing the stiffeners within a range of radial locations approximately varying from 70 mm to 110 mm. This is an important inference and we will use it for the further analysis on the dynamic LFD.

Based on the above analysis, the helium vessel stiffness was increased by increasing its wall thickness from 4 mm to 5 mm [24], and increasing the internal torus radius of the end cover from 35 mm to 120 mm. In order to facilitate the joining between niobium cavity and titanium vessel, in this model we also accommodate a transition piece of 55Ti-45Nb as shown in Fig. 22.

The above analysis on the static LFD gives us ideas about the stiffness requirement of the assembly, the preferred radial locations of the stiffener rings and the tuning range required to compensate the static LFD. We now discuss the transient analysis by analyzing the pressure pulse in time as well frequency domain. The temporal shape of the input voltage pulse is shown in Fig. 23. The $H^-$ ion beam is injected for a flat-top duration of 2 ms, with a rep. rate of 50 Hz. The rising and falling part of the RF pulse shape is determined by the cavity loaded quality factor of the cavity. As a consequence of this periodic RF pulse, the pressure pulse will also be periodic and will repeat with the same PRR. Figure 24 is the frequency domain representation of the pressure pulse. The resonant forced oscillations will appear if the frequency of the mechanical mode of the cavity is close to the frequency of one of the Fourier components of the pressure pulse. If this happens, the stiffener position should be modified such that this equality condition is avoided. Figure 24 indicates that in the pressure pulse, relative amplitude for frequencies higher than 250 Hz is significantly small. This means that for frequencies greater than 250 Hz, even if resonance condition is satisfied, it will cause small dynamic displacements.

We now present the structural mode frequencies of the modified cavity-helium vessel assembly with stiffener rings. We have performed simulations for different radial locations of the stiffener rings,by varying its mean position from 108 mm to 124 mm. The results are in Table VI. According to the data shown in Table VI, we conclude that, though the the third participating mode produced because of the stiffener radial position of 116 mm, none of the structural mode frequencies are multiples of 50 Hz, which is the PRR of the RF. This tells us that if we put the stiffener rings at the radial locations other than 116 mm, we can expect that there will be a little amplification because of the dynamic LFD. However, particularly, the stiffener location at 124 mm will be the most suitable choice because in that case, the first structural mode is at 265.07 Hz, and is greater than 250 Hz, which ensures that all the structural mode frequencies will have progressively lower contributions in the total response[24]. By moving the stiffener radial position further in the outward direction, we can expect that the first as well as the other modes will move far away from 250 Hz. But as we observed during the static LFD analysis, if we move away from the radial location range of 70 mm to 110 mm, tuning becomes difficult. Fig. 25 shows that the argument is still valid for the modified cavity-helium vessel assembly. This plot shows that as we move radial position of the stiffener rings upward, a larger displacement will be required to compensate the LFD. Therefore, the radial location of the stiffener rings at 124 mm is the optimum choice and in this radial position of the stiffener rings, in order to compensate for the LFD, the helium vessel needs to be displaced by 7.35 μm.

**VI. Discussions and Conclusions**

In this paper, we have presented the details of calculations performed for the electromagnetic design of a $\beta_g$= 0.61,650-MHz elliptic multi-cell SRF cavity. We have arrived at an optimized cavity geometry following a step by step, one dimensional optimization procedure developed earlier in Ref. 2 for $\beta_g$= 0.9 cavity. There are however several new features in the procedure that we have added in the work presented in this paper. We would like to emphasize these in this section.

We have described a procedure for choosing the iris radius. As we increase the iris radius, the inter-cell coupling coefficient and the group velocity increases. This has an important implication. If a particular cell has a resonant frequency shifted from the operating RF frequency, there is a reflection from the cell at the location of the iris. The amplitude of reflected wave is inversely proportional to the inter-cell coupling coefficient. The reflected wave from different cells interfere and modulated the electric field amplitude in different cells. This deteriorates the field flatness. Therefore, we choose the iris radius to ensure that there is enough inter-cell coupling such that, including the cell to cell frequency error arising due to limitation in manufacturing and processing, the desired field flatness is achieved.

For the design of end cells, we have described a procedure to tune the cavity geometry to avoid trapped monopole HOMs. For this, the independent variables describing the end cell geometry are tuned such that it resonates at the same frequency as mid cells, both for operating mode as well as monopole HOM that is otherwise trapped.

As already emphasized in the paper, for $\beta_g = 0.61$ cavity, the calculation of threshold current for regenerative BBU is done for a range of values of $\beta$. This is a general feature, while designing mid $\beta$ cavities. The reason is that in mid $\beta$ cavities, there is a wider range of particle velocity, compared to high $\beta$ cavities. Therefore, a more elaborate calculation of threshold current needs to be performed, which we have done in this paper, and we have outlined the procedure for this calculation. Another special feature while designing mid $\beta$ cavities is that for the wakefield calculation, the ultra-relativistic approximation that is usually implied in calculations for high $\beta$ cavities, is not valid. We have given a simple procedure to evaluate the longitudinal loss factor for mid $\beta$ cavities, and verified that our approach reproduces the predicted results by standard code, when applied for $\beta = 1$ case.

Finally, we have discussed the procedure for performing LFD calculations to optimize the stiffening design of the cavity. For pulsed operation, the dynamic LFD is important. Dynamic LFD calculations are more involved and time consuming. Static LFD calculations are relatively simple and give a good indication of dynamic LFD behavior. It is therefore recommended that static LFD calculations are first performed and the calculations are fine-tuned based on dynamic LFD analysis. In this paper, rather that performing a complete transient analysis, we have analyzed the structural modes of the cavity along with its stiffening mechanism and ensure that it has no strong Fourier component at frequency close to the frequency component present in the pressure pulse generated by Lorentz force due to RF pulse.

We would like to briefly discuss the contemporary design studies for $\beta_g$ = 0.61,elliptical SRF cavity available in the literature. For the SNS project at oak-ridge[1], optimization of 6-cell, $\beta_g$ = 0.61, 805 MHz, elliptic SRF cavities were performed to fulfill their requirement of accelerating an average proton beam current ~ 1 mA at the target. Another contemporary activity is the design and optimization of $\beta_g$ = 0.61, 650 MHz SRF elliptical cavities for Project-X [25], where optimization is done to maximizethe accelerating gradient[25]. In addition to this, European Spallation Source[ESS] project is planning to use $\beta$ 0.67 SRF cavities for accelerating the charged particles in the medium energy range [26].Compared to the earlier publications, in this paper, we have presented a generalized approach of designing an SRF elliptical cavity that addresses the interlinked design issues in a systematic manner. Starting from the electromagnetic design optimization of the cavity geometry, we have presented a detailed analysis of HOMs, wakefields and LFD. We can summarize the important aspects of the optimization procedure discussed in the paper as follows:

1) First, we arrive at an optimized mid cell geometry following a step by step, one dimensional optimization procedure.

2) To obtain the preliminary design of the cavity, we adjust the end half-cell length in order to restore the cavity resonant frequency.

3) If there is a trapped monopole mode, we tune the end half-cell length $L_e$ and major equator ellipse radius $A$ together such that the end cell resonates at the same frequency as that of multi-cell cavity for the fundamental as well as higher order mode. This ensures that the HOM is not trapped.

4) Following the rigorous analysis of theregenerative beam break up (BBU) instability presented in the paper, the threshold current for the BBU instability should be calculated for all prominent dipole modes for the range of beam velocities for which the cavities will be operated. It should be ensured that the operating beam current is lower than this current.

5) Following the approach discussed in the paper, the optimum position of the stiffener ringsto ensure that the cavity resonance frequency can be tuned back to the design value within its bandwidth, even in the presence of dynamic LFD.

To conclude, we have presented a detailed electromagnetic design of a $\beta_g = 0.61$,650-MHz elliptic multi-cell SRF cavity for the ISNS project, where the cavity geometry has been optimized to maximize the acceleration gradient, and effect of HOMs and LFD have been included in the analysis. A generalized procedure for these calculations is evolved, which is elaborated in this paper. Detailed beam dynamics study through these cavities along with suitable focusing lattice will be taken up later.

**Acknowledgements**

The authors are grateful to Mr. Abhay Kumar for his valuable ideas and help for the calculation of the LFD and help in ANSYS simulations in High Frequency Domain that he extended to. Also, it is a pleasure to thank Dr. P. D. Gupta for constant encouragement, and Dr. S. B. Roy for helpful discussions and suggestions on the manuscript.

**Figures:**

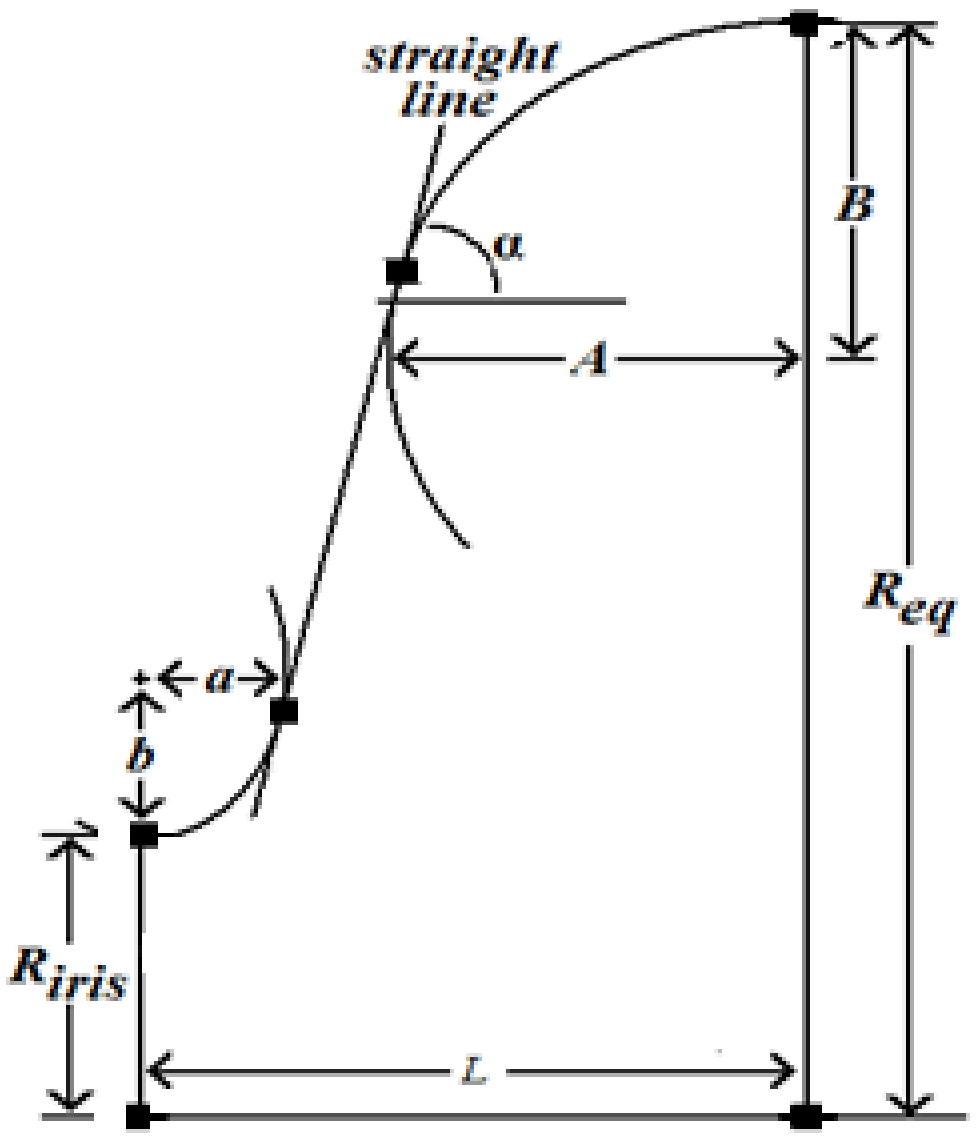

Fig. 1. Schematic of the half-cell of an elliptic cavity.

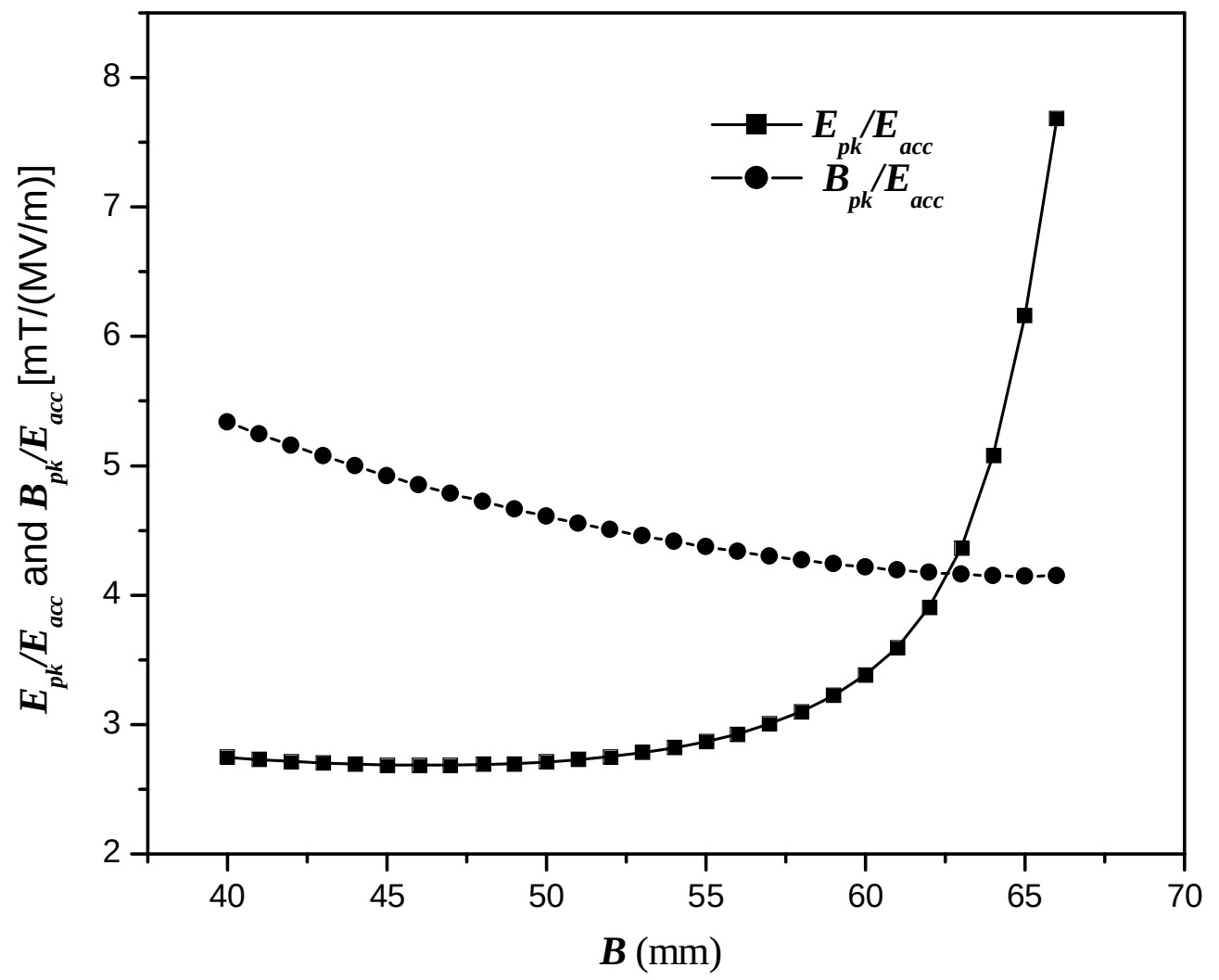

Fig. 2.Variation of $B_{pk}/E_{acc}$ and $E_{pk}/E_{acc}$ as a function of $B$. Here, $a/b = A/B = 1$, and $\alpha = 88^0$.

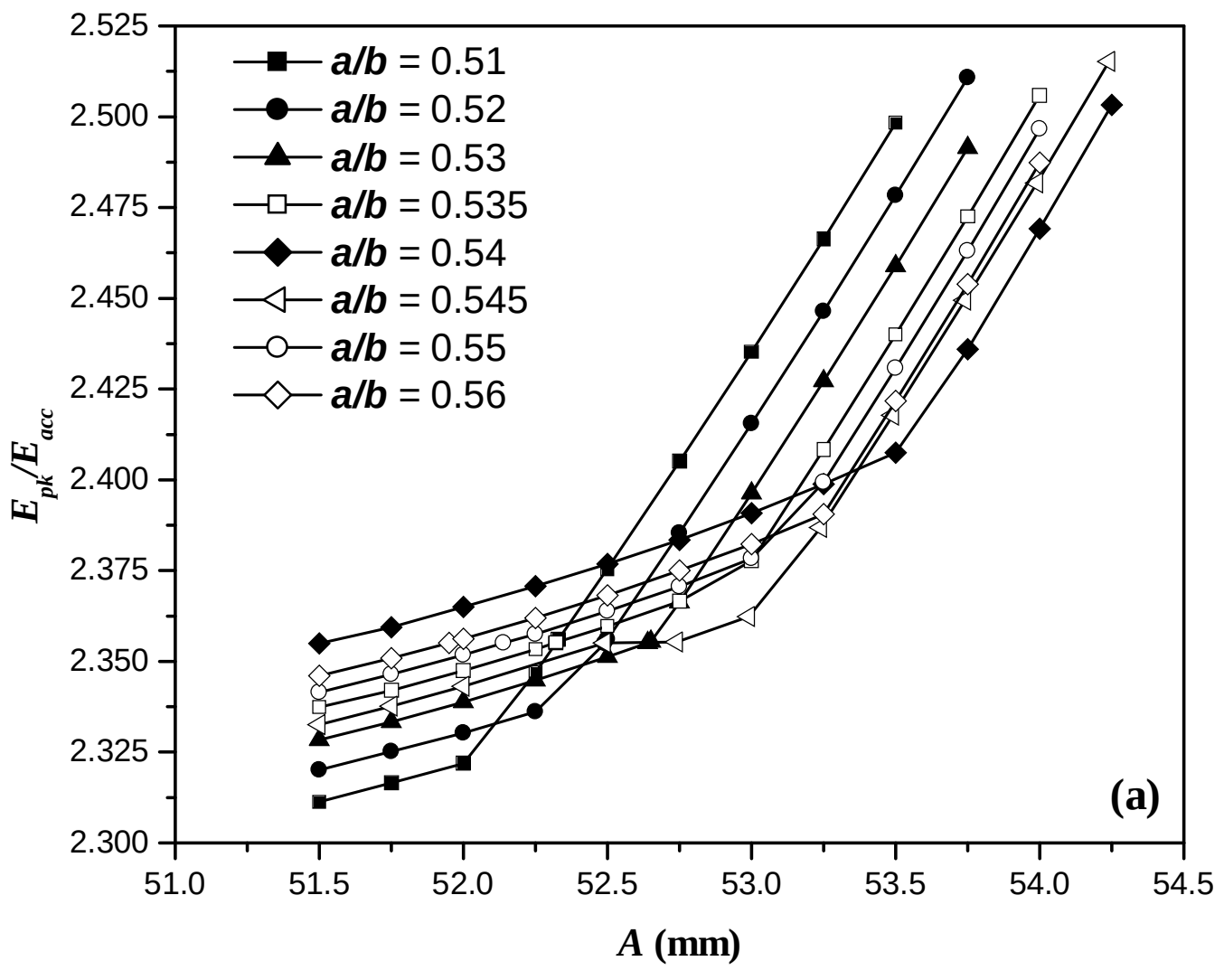

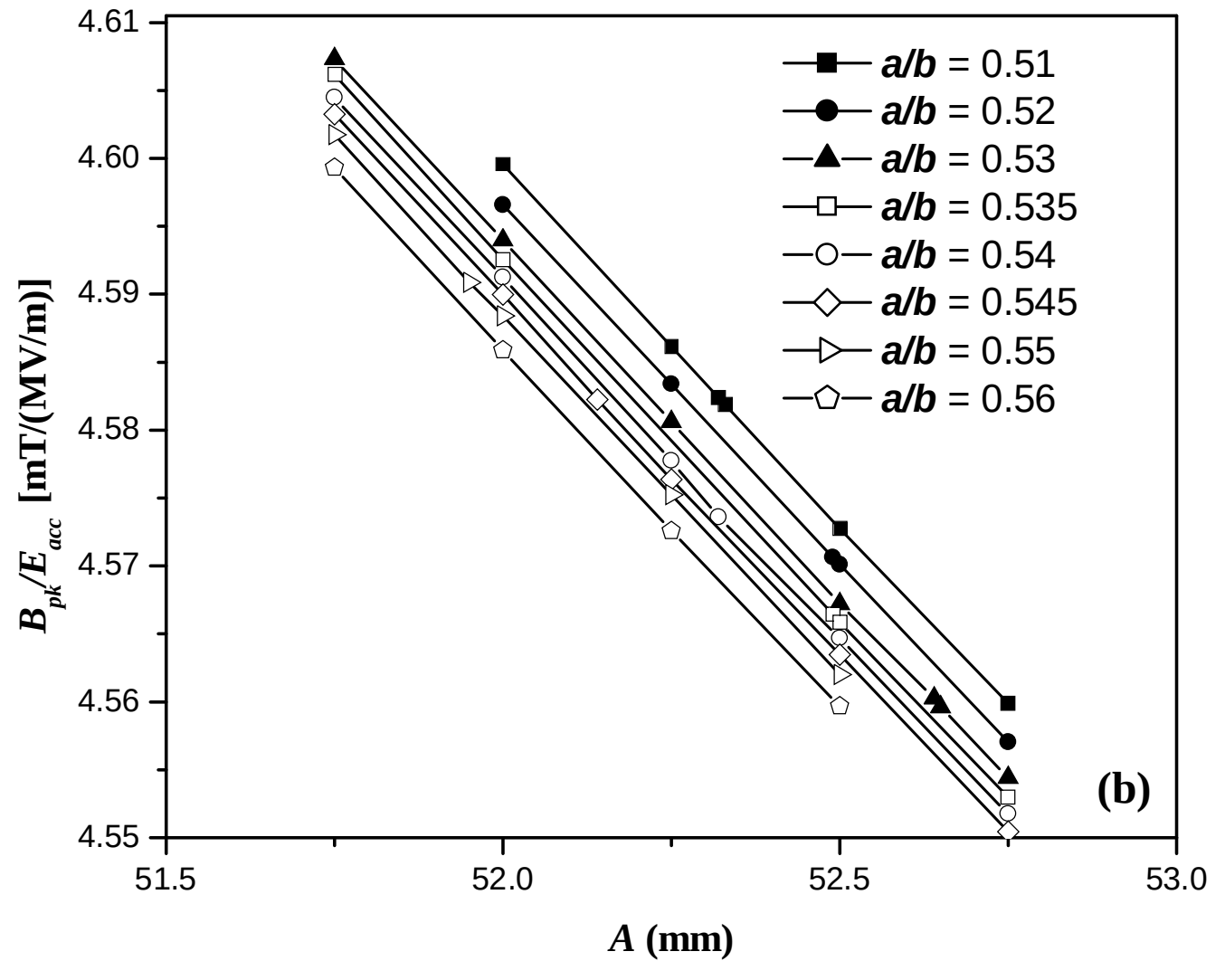

Fig. 3. Plot of (a) $E_{pk}/E_{acc}$ as a function of $A$, and (b) $B_{pk}/E_{acc}$ as a function of $A$.

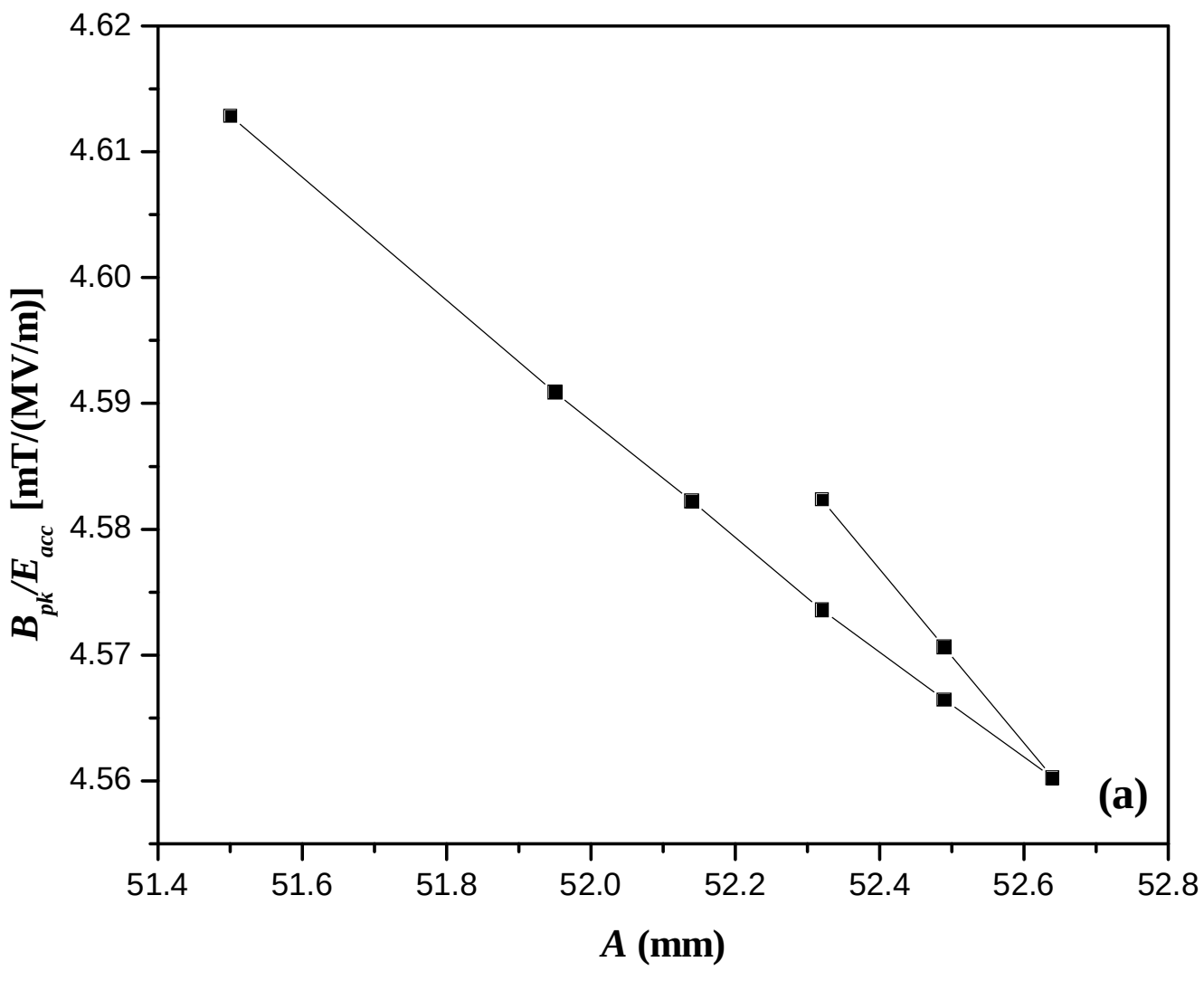

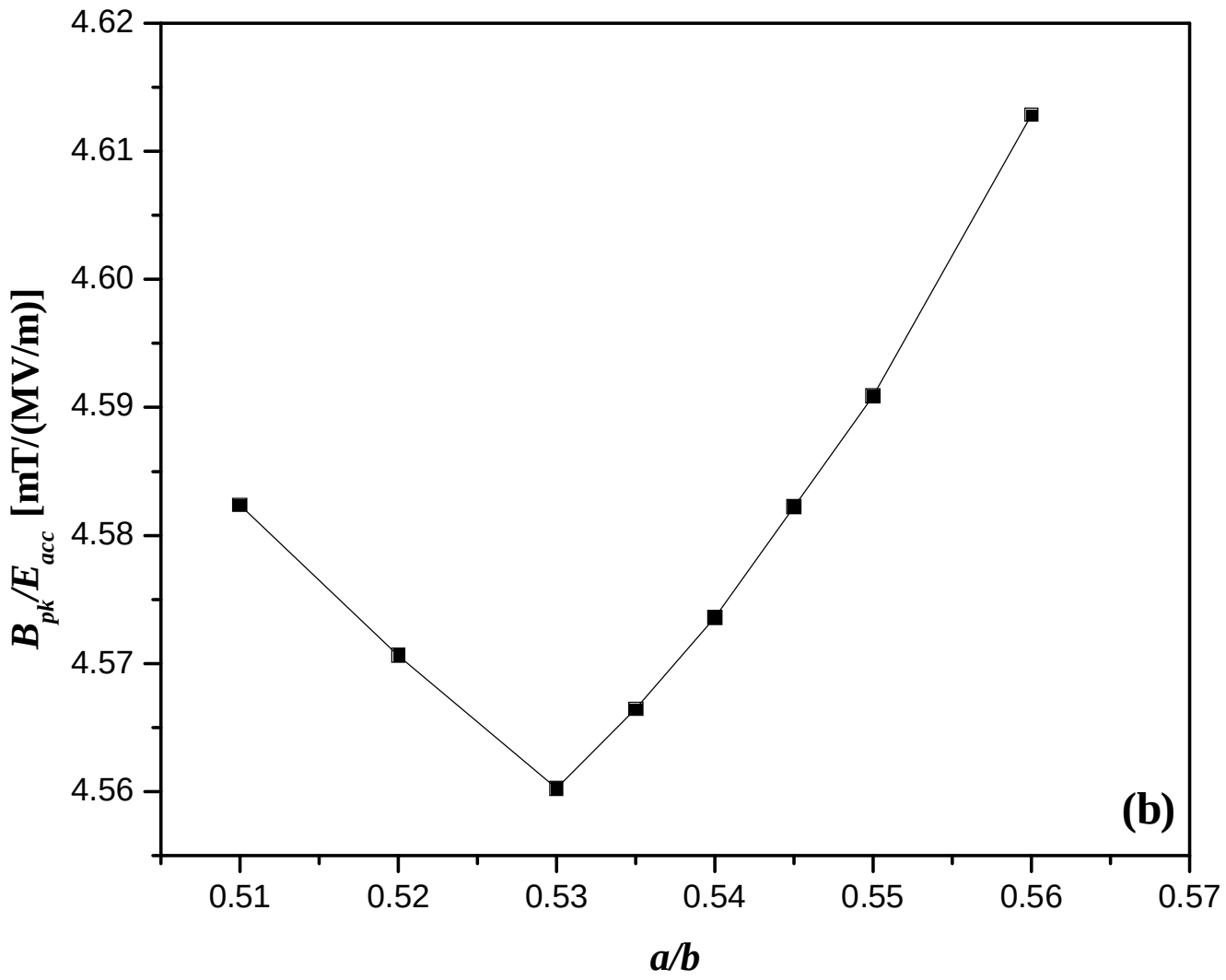

Fig. 4. Plot of $B_{pk}/E_{acc}$ (a) as a function of *a/b* and (b) as a function of *A*. For each of these points, $E_{pk}/E_{acc}$= 2.355. Data for plotting these curves are taken from Fig. 3.

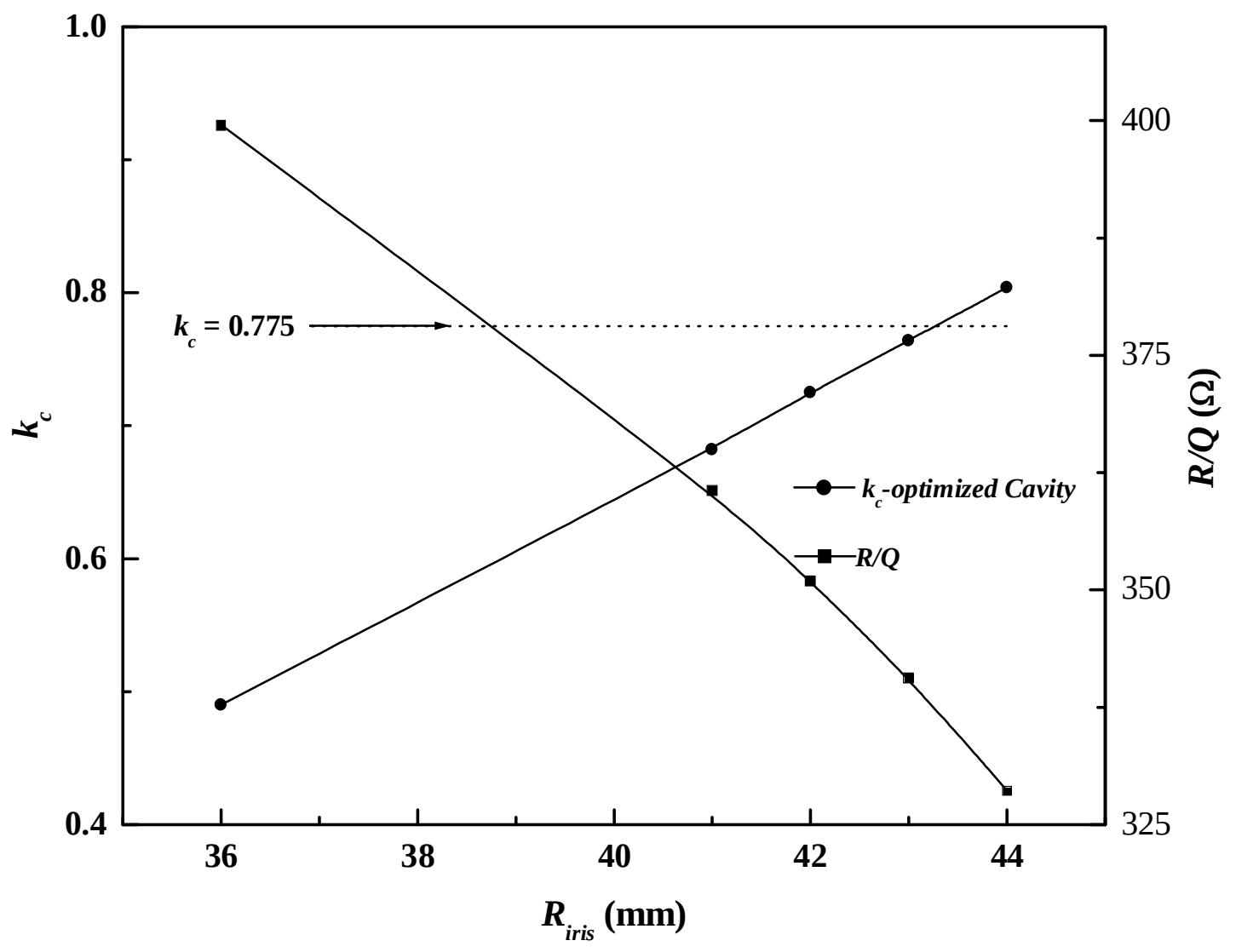

Fig. 5. Plot of $k_c$ (and *R/Q*) with the $R_{iris}$. Here the dotted line shows the value of $k_c$ ~ 0.775.

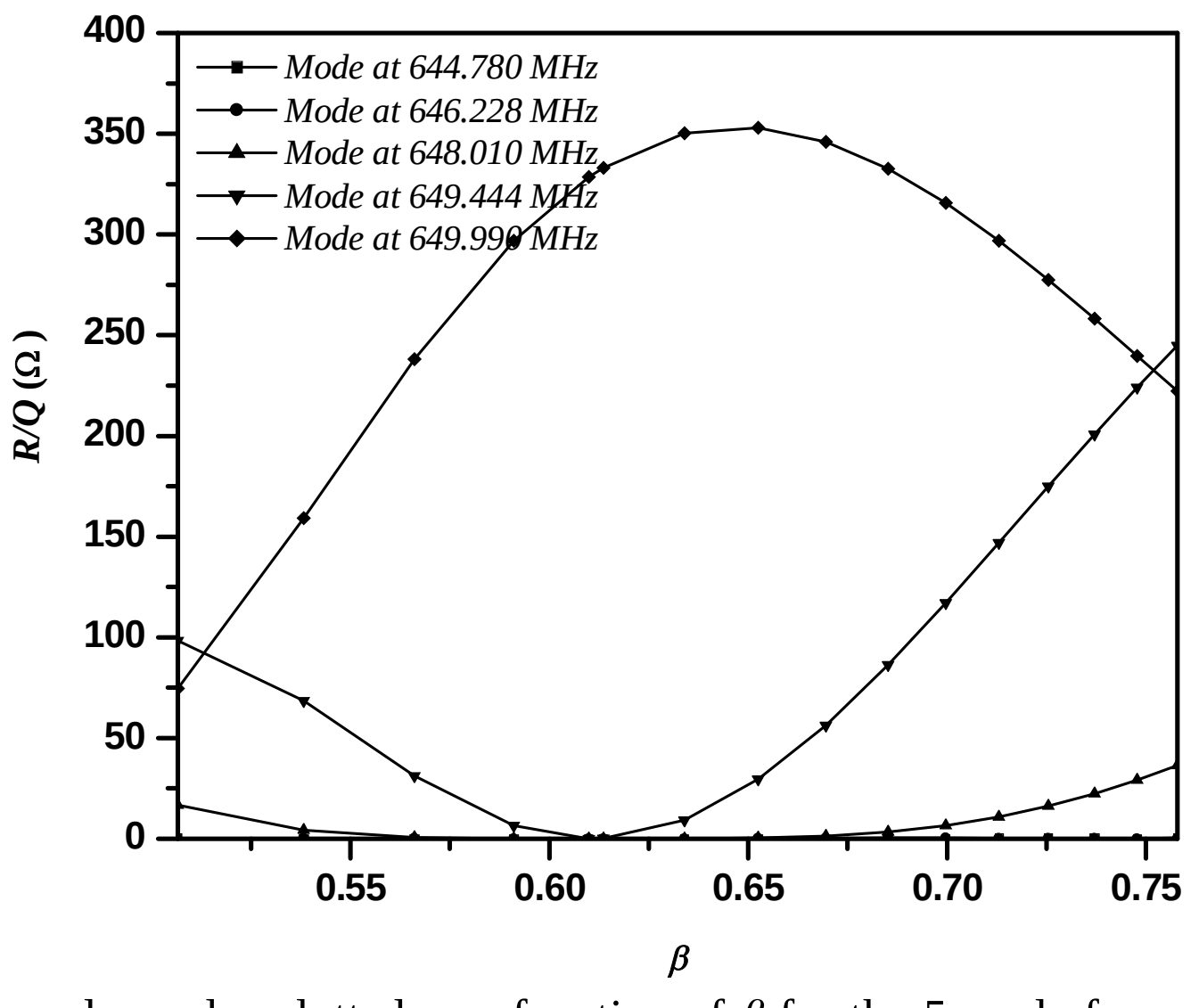

Fig. 6. *R*/*Q* of the monopole modes plotted as a function of *β* for the 5 mode frequencies of the first pass band.

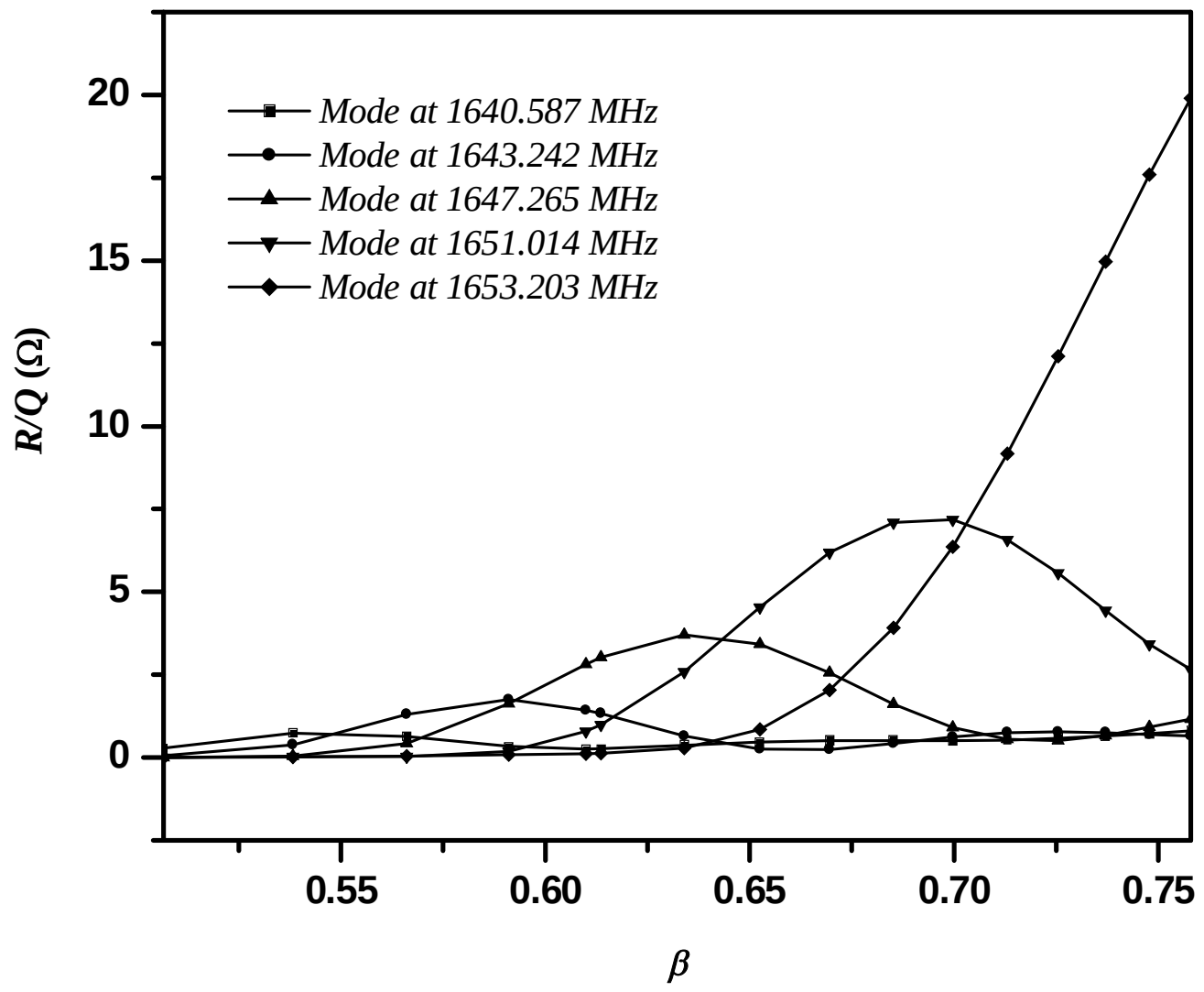

Fig. 7. $R/Q$ of the monopole modes plotted against $\beta$ for the five mode of the third pass band.

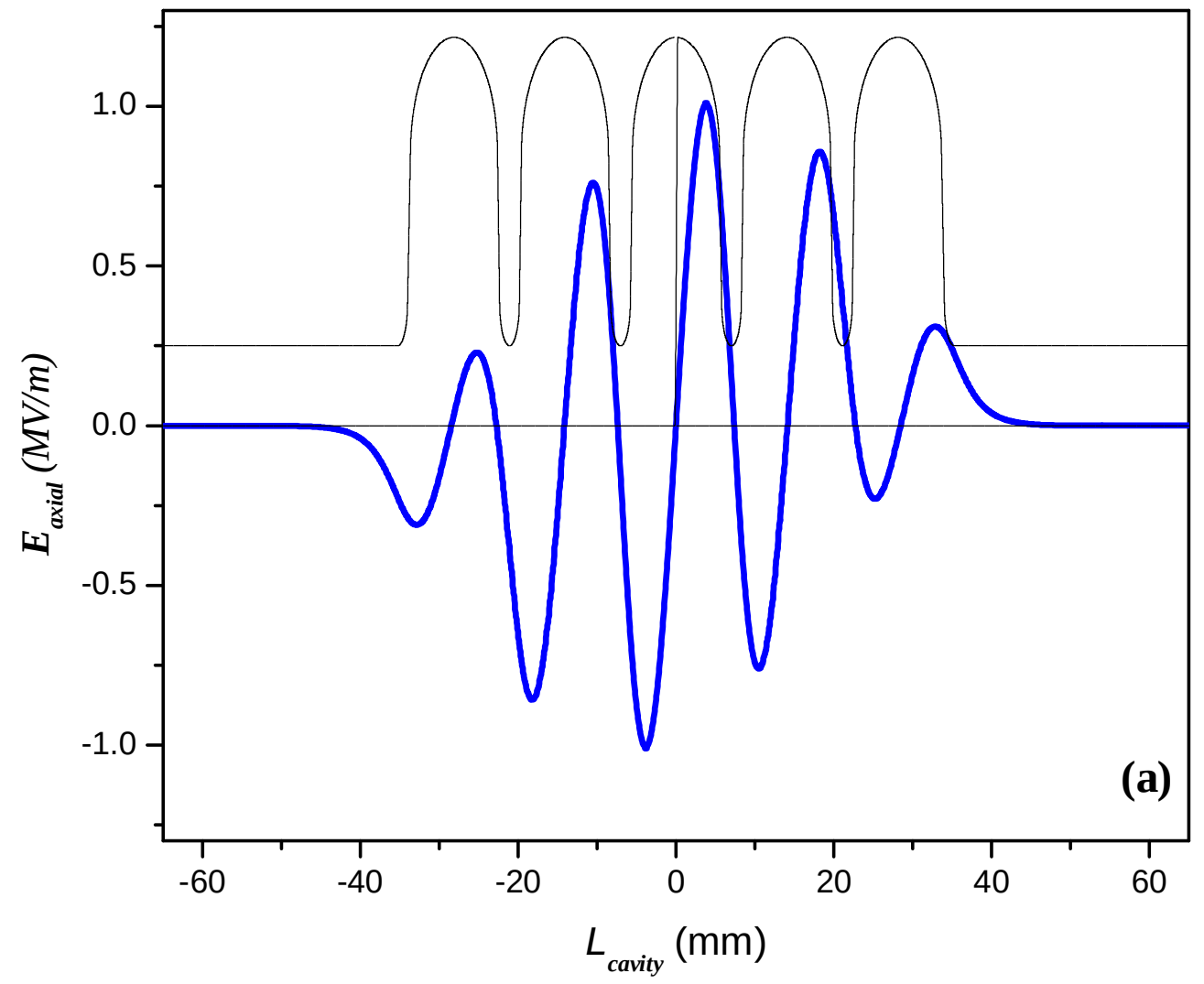

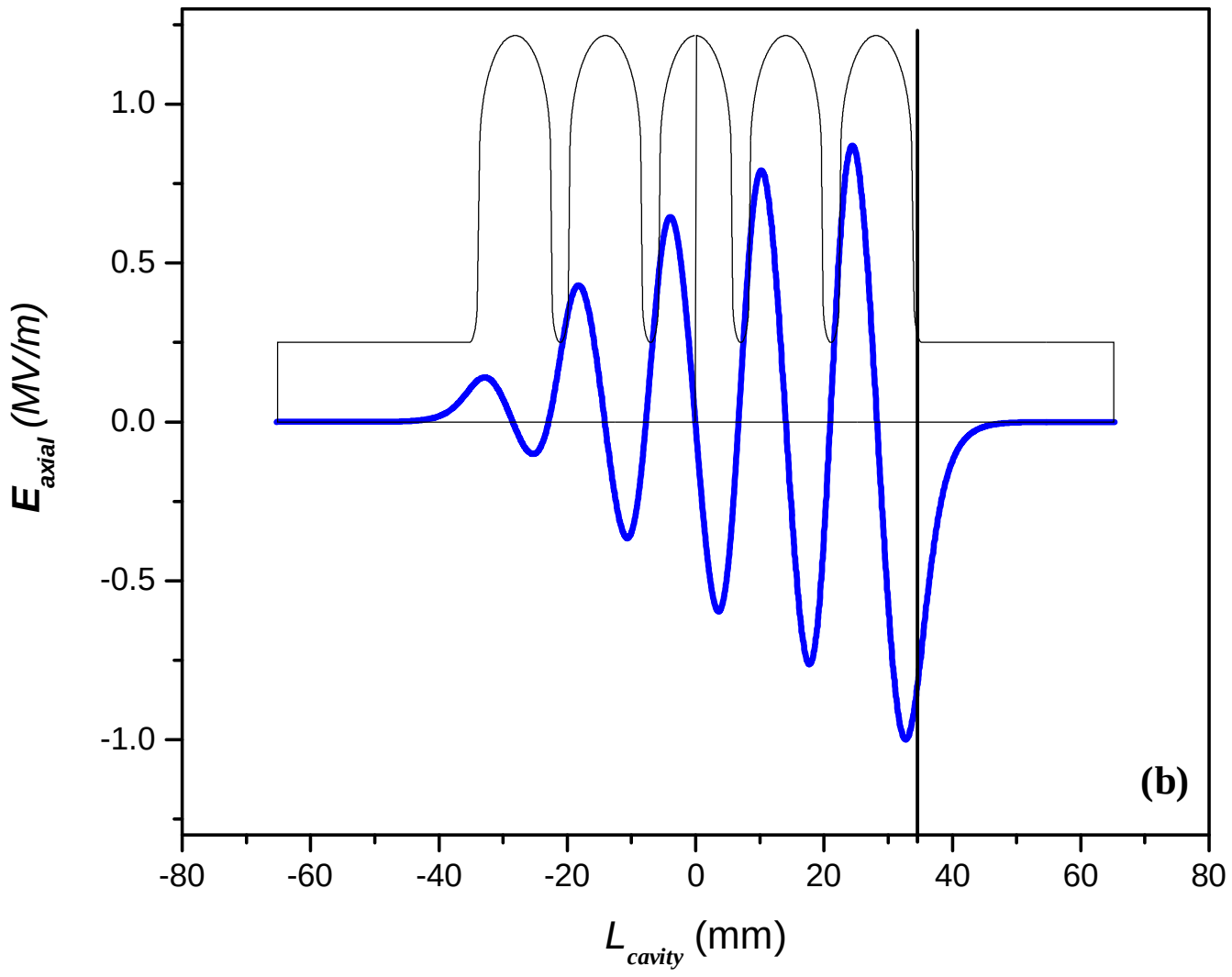

Fig. 8. Axial electric field of the mode at 1653.20 MHz is plotted with the cavity length. Here, cavity midpoint coincides with the origin. (a) This is the mode configuration for the unmodified geometry. Mode trapping is clearly seen inside the cavity. (b) This is the mode configuration for the geometry with modified end cell at one side.It is seen that the field is no more trapped inside tha cavity.

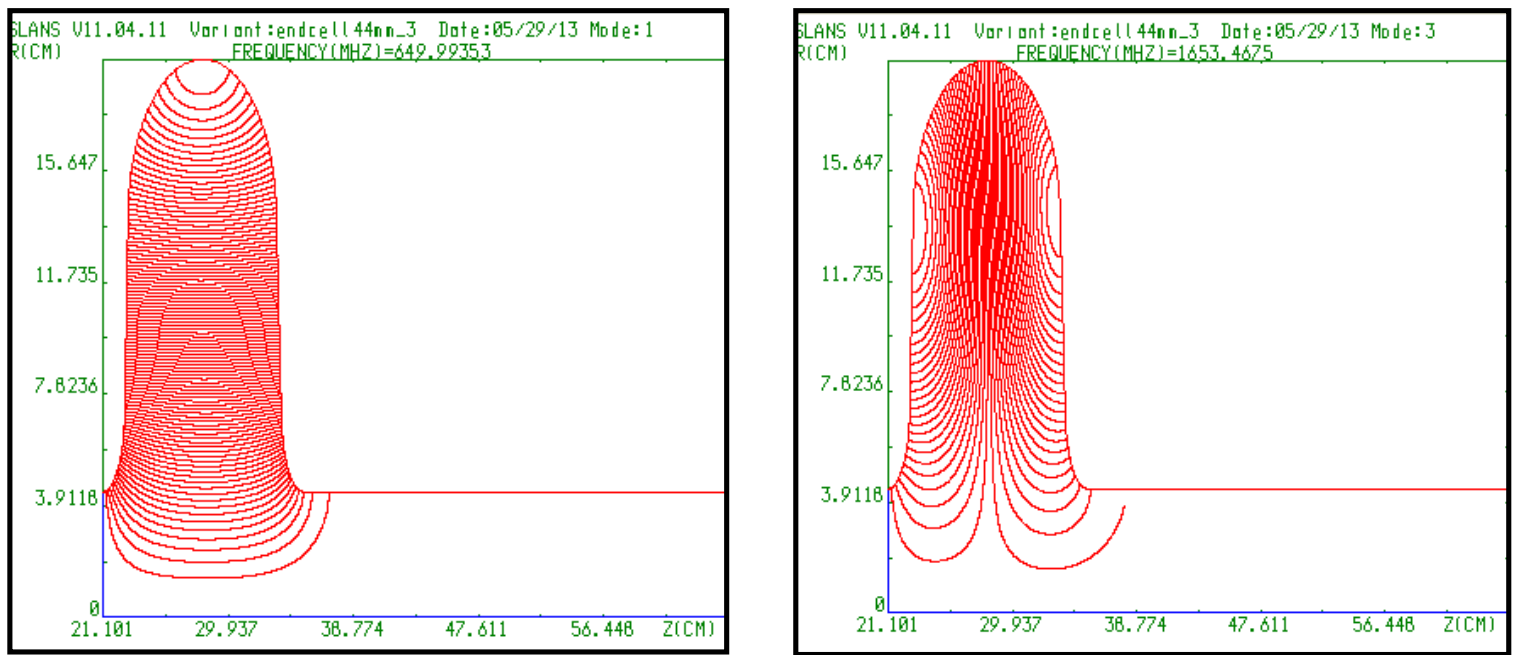

Fig. 9. The modified end cell will act as a resonator for the fundamental mode as well as for the particular HOM. (a) the field contour for the fundamental mode. Here the end cell resonates at 650.0 MHz. (b) The same end cell also acts as a resonator for the HOM at 1653.20 MHz. Electric field contours are shown in the figure.

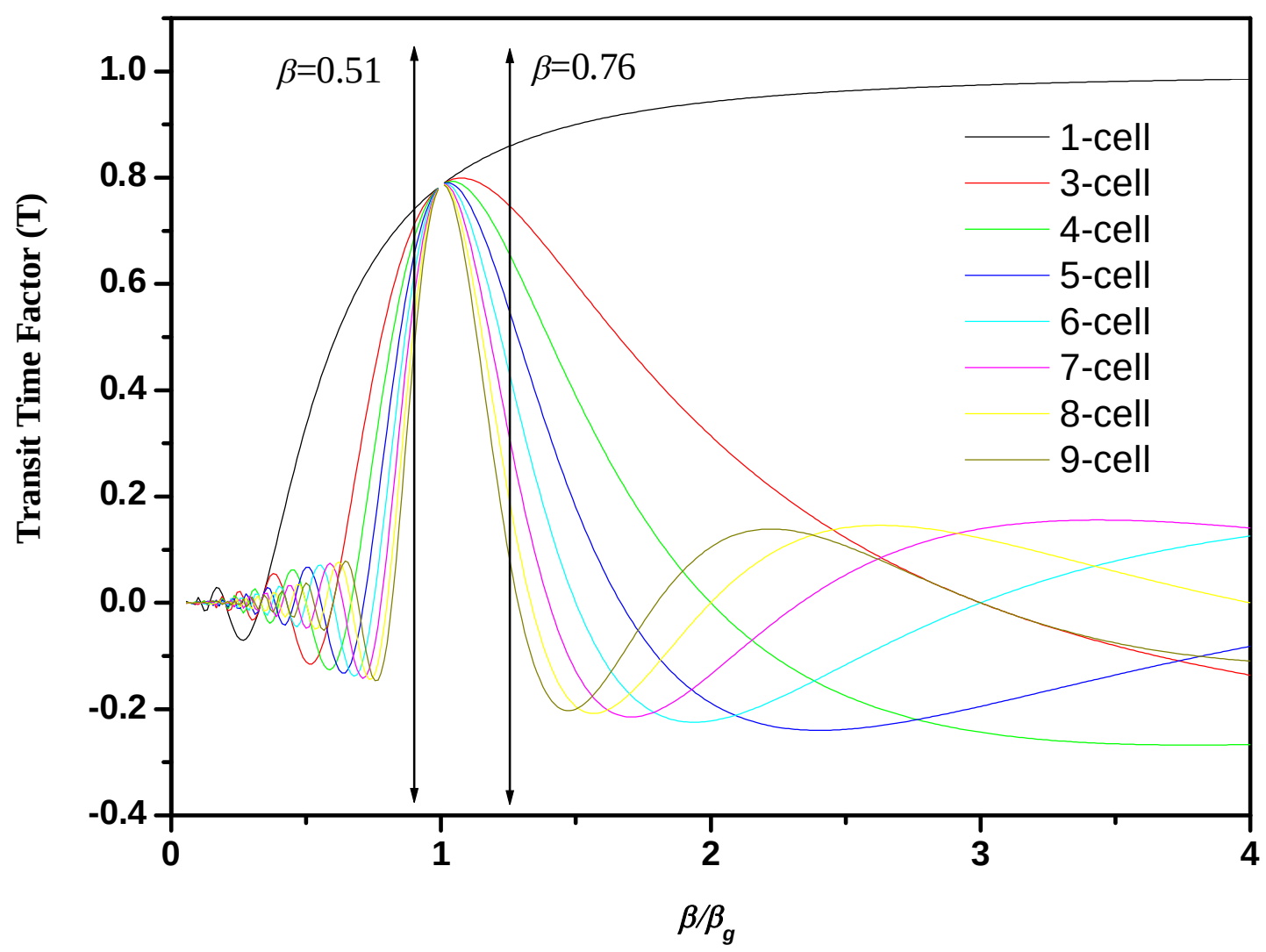

Fig. 10. Variation of the transit time factor $T$ as a function of the normalized particle velocity ($\beta/\beta_g$). The two vertical lines correspond to $\beta/\beta_g$ = 0.85 and 1.27, as explained in the text.

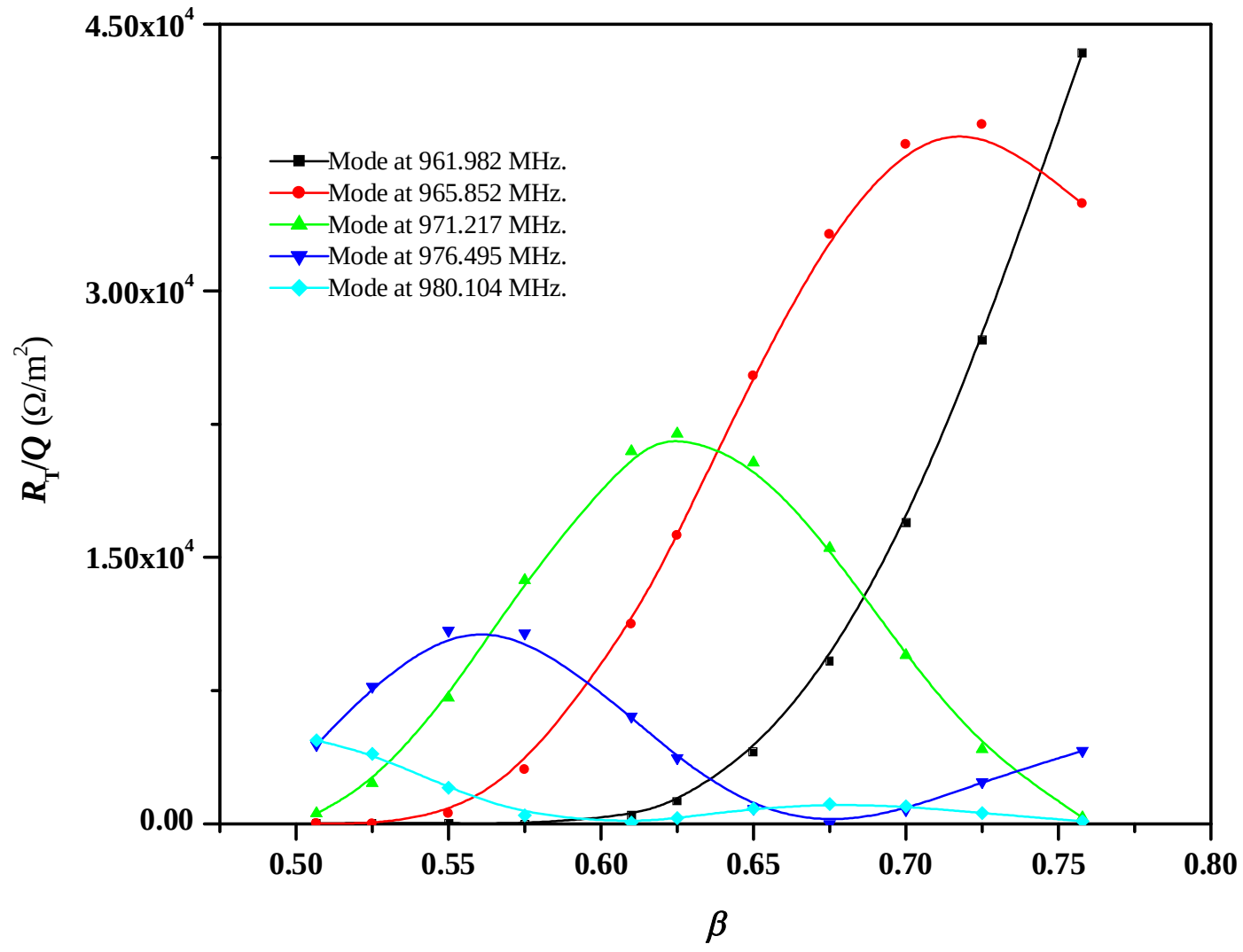

Fig. 11.$R_\perp/Q$ of the dipole modes plotted as a function of $\beta$ for the five modes of the first pass band.

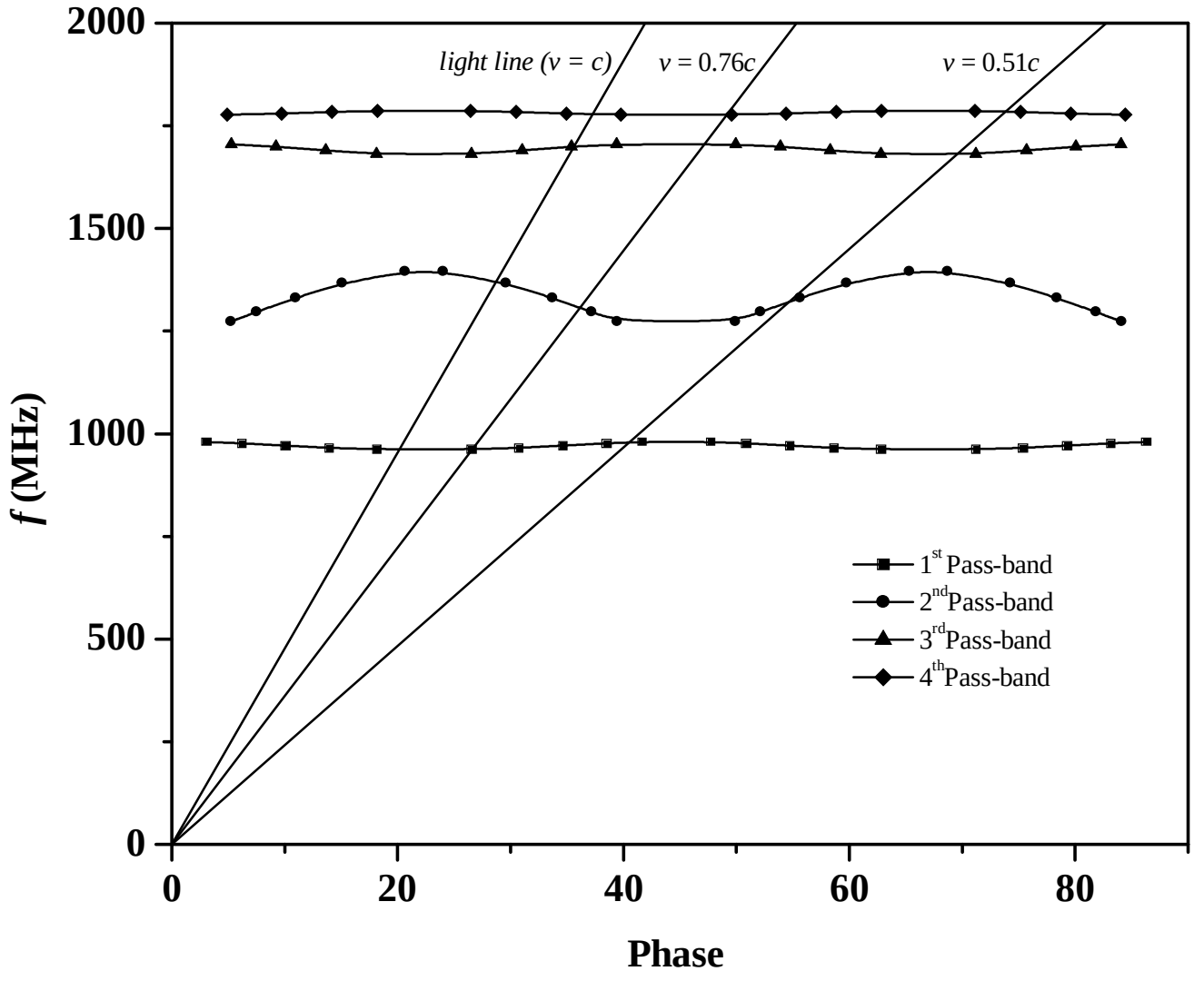

Fig. 12. Dispersion diagram for the first few dipole pass bands. Here mode frequencies are plotted against their corresponding phase advance per cell.

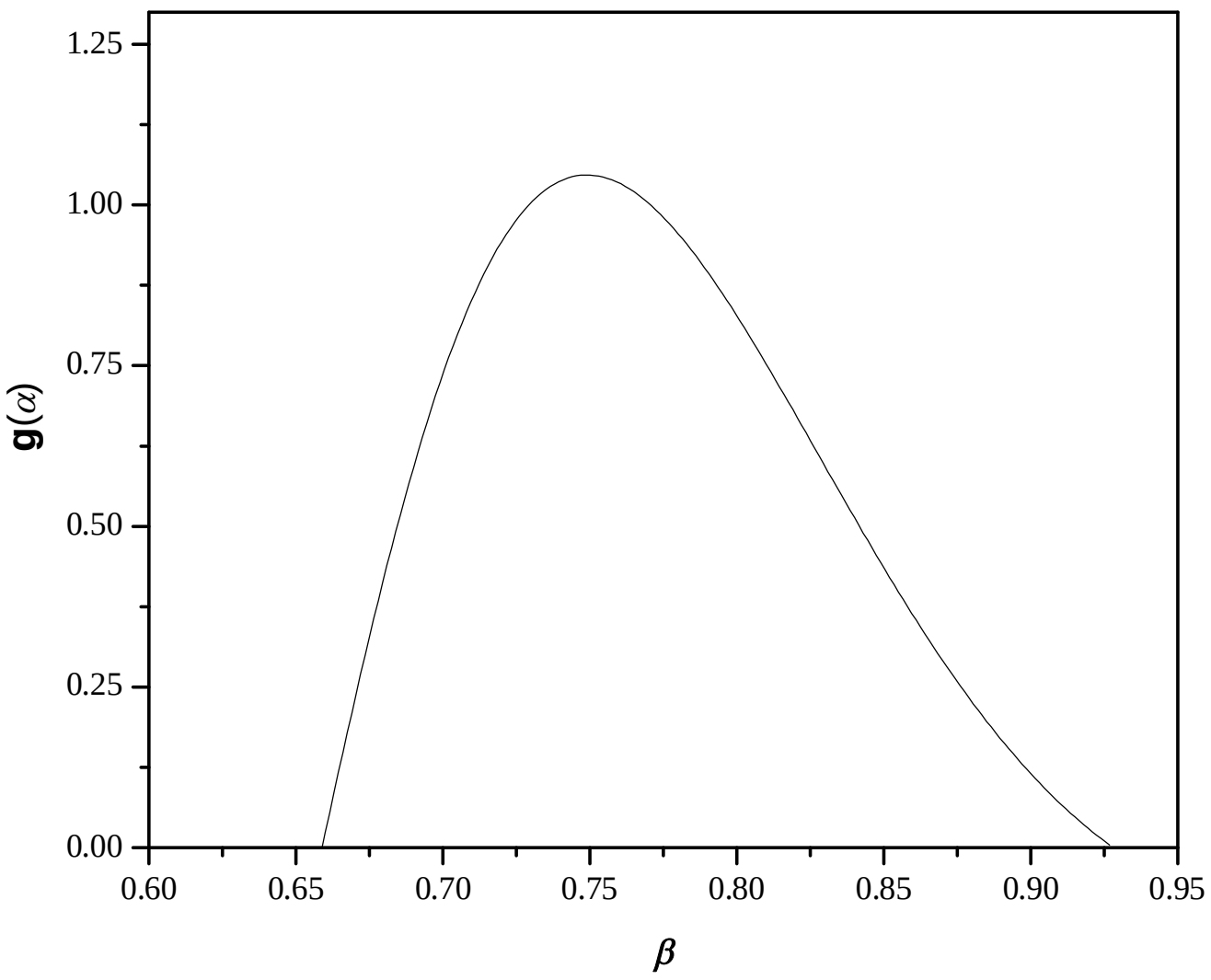

Fig. 13.Dependence of $g(\alpha)$ on$\beta$ for the mode resonating at 965.852 MHz.

.

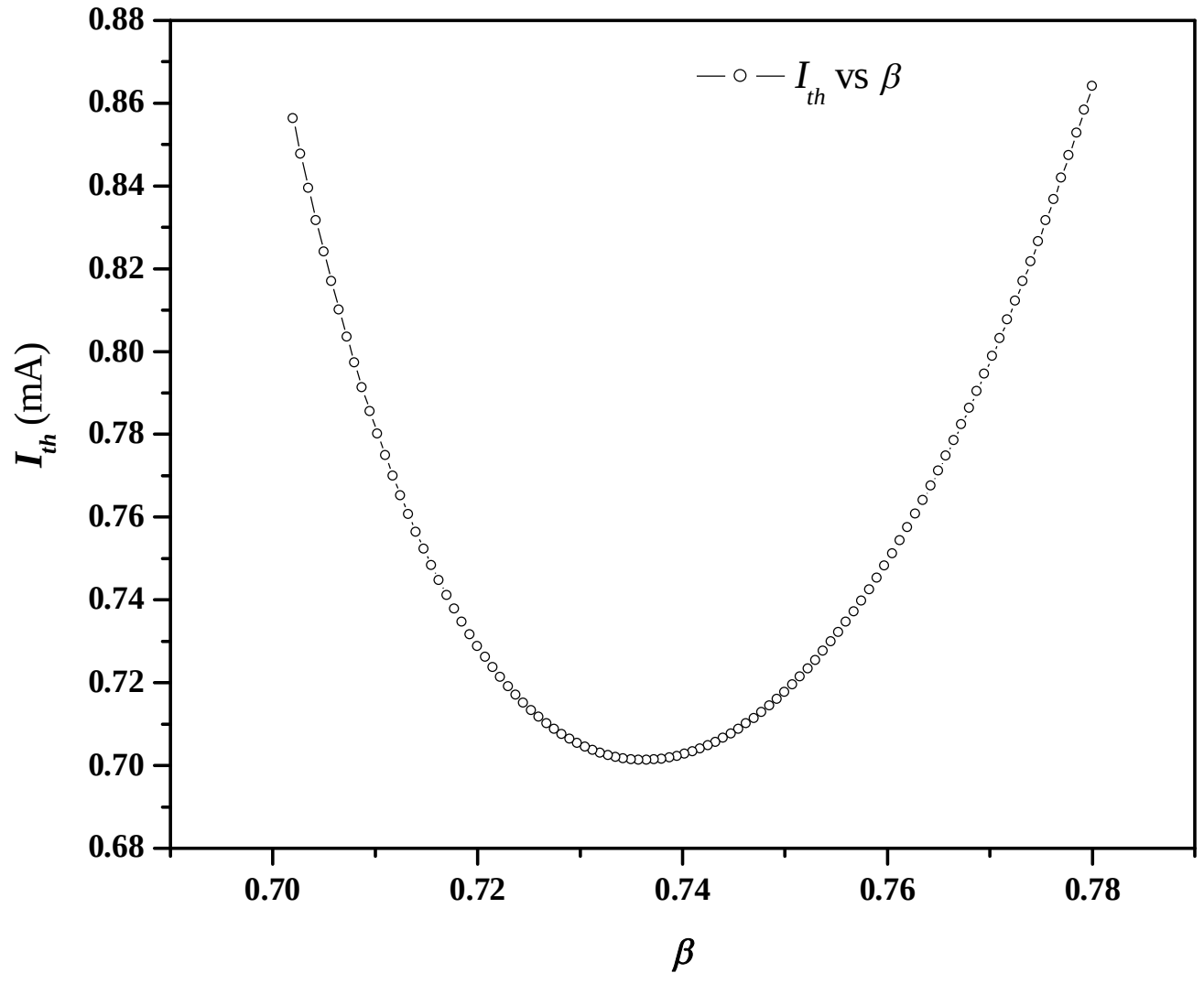

Fig. 14.Dependency between $I_{th}$ and $\beta$ for the mode resonating at 965.852 MHz.

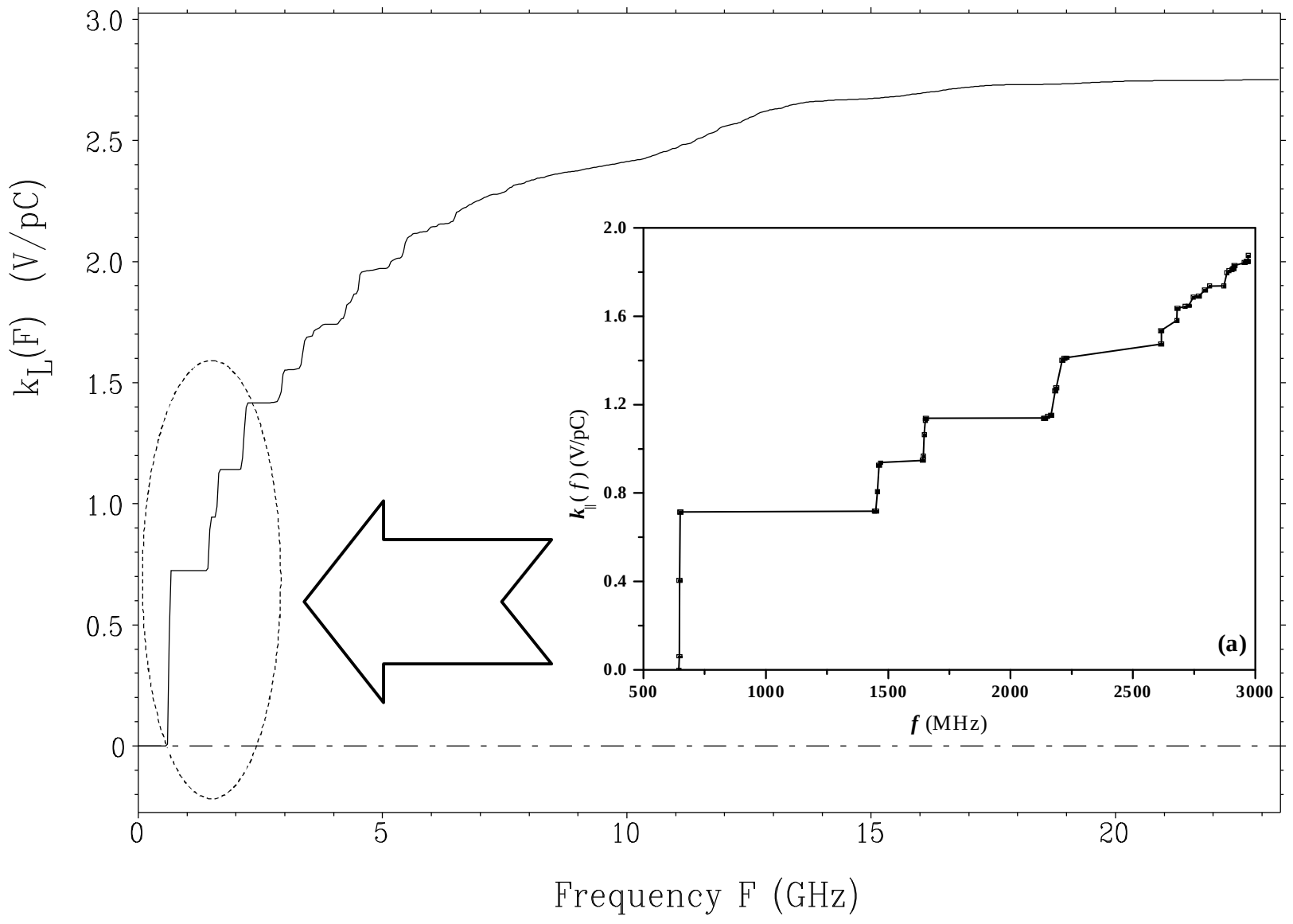

Fig. 15. Longitudinal loss factor $\kappa_{\|}$ (or $k_L$) as a function of the frequency $f$. (a) Values calculated analytically for $\beta\rightarrow 1$and (b) the same obtained from ABCI for a Gaussian bunch of $\sigma$ ~ 5 mm.

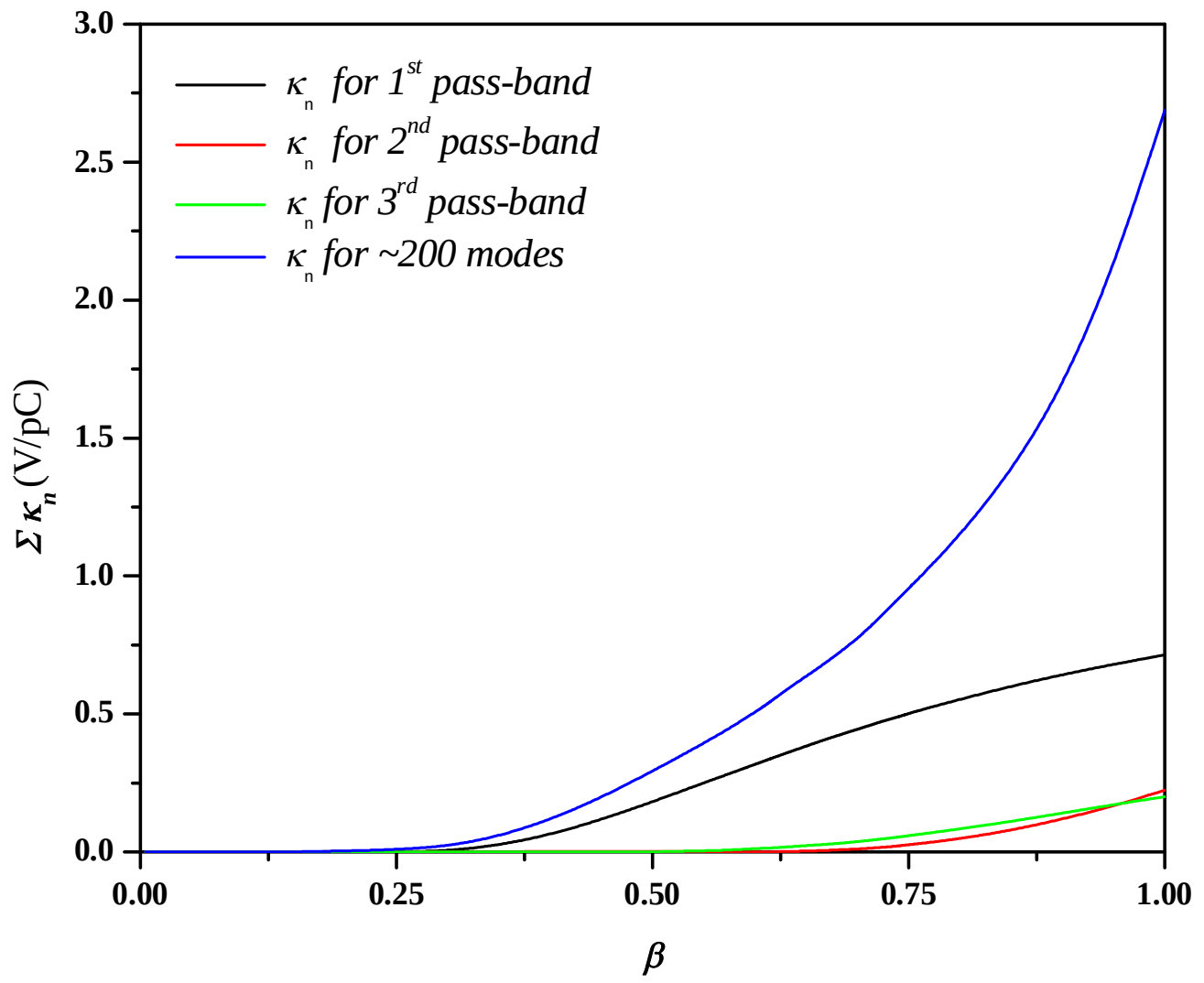

Fig. 16. Here integrated loss factor $\Sigma\kappa_n$ is plotted with $\beta$. The blue lineis generated by summing over the longitudinal loss factors of ~ 200 modes.

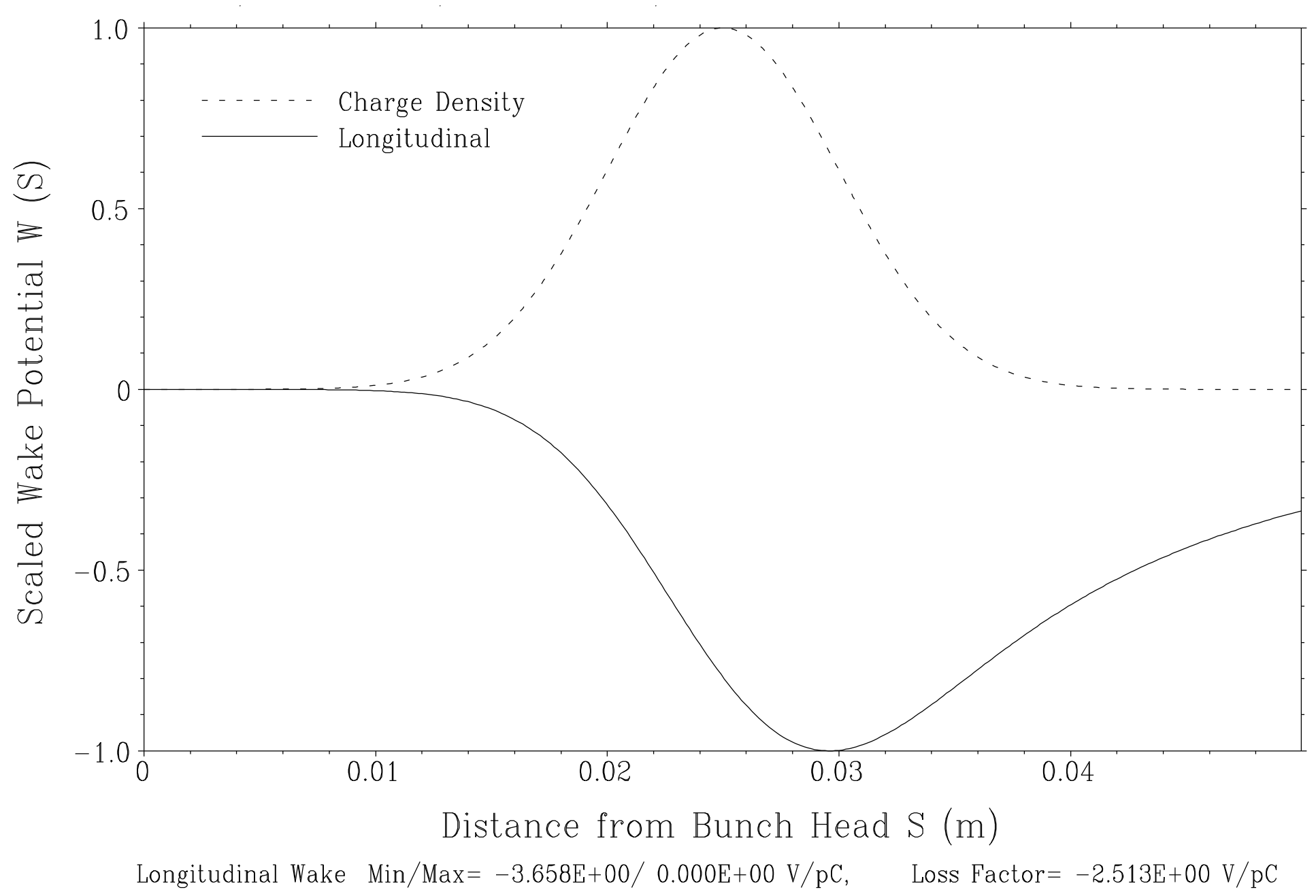

Fig. 17.Longitudinal and transverse wake potential generated by a Gaussian bunch having 1 pC charge passing through the optimized 5-cell cavity.

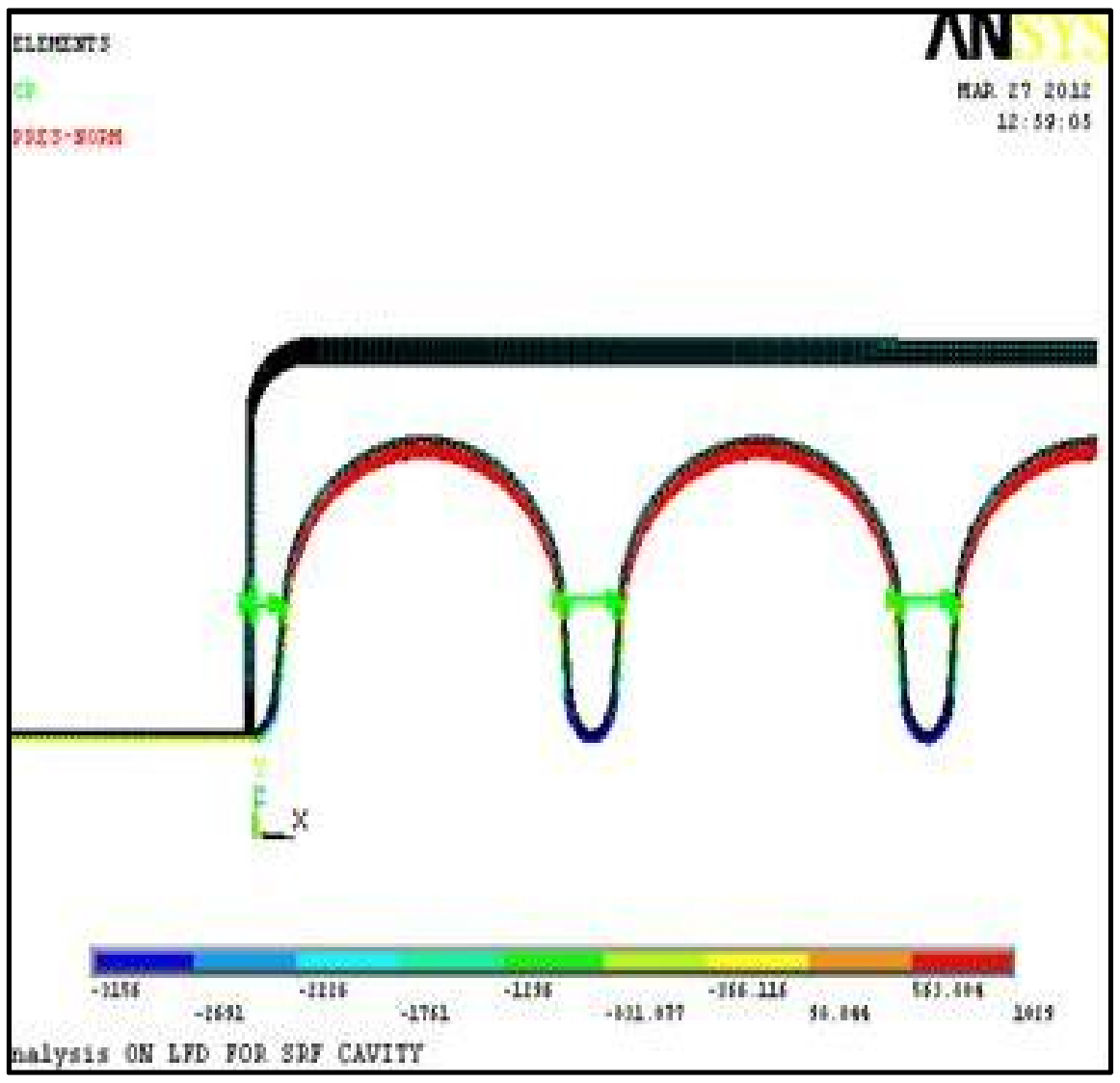

Fig. 18.Schematic of the multi-cell cavity along with the helium vessel. Green lines connecting the cells are the stiffeners.

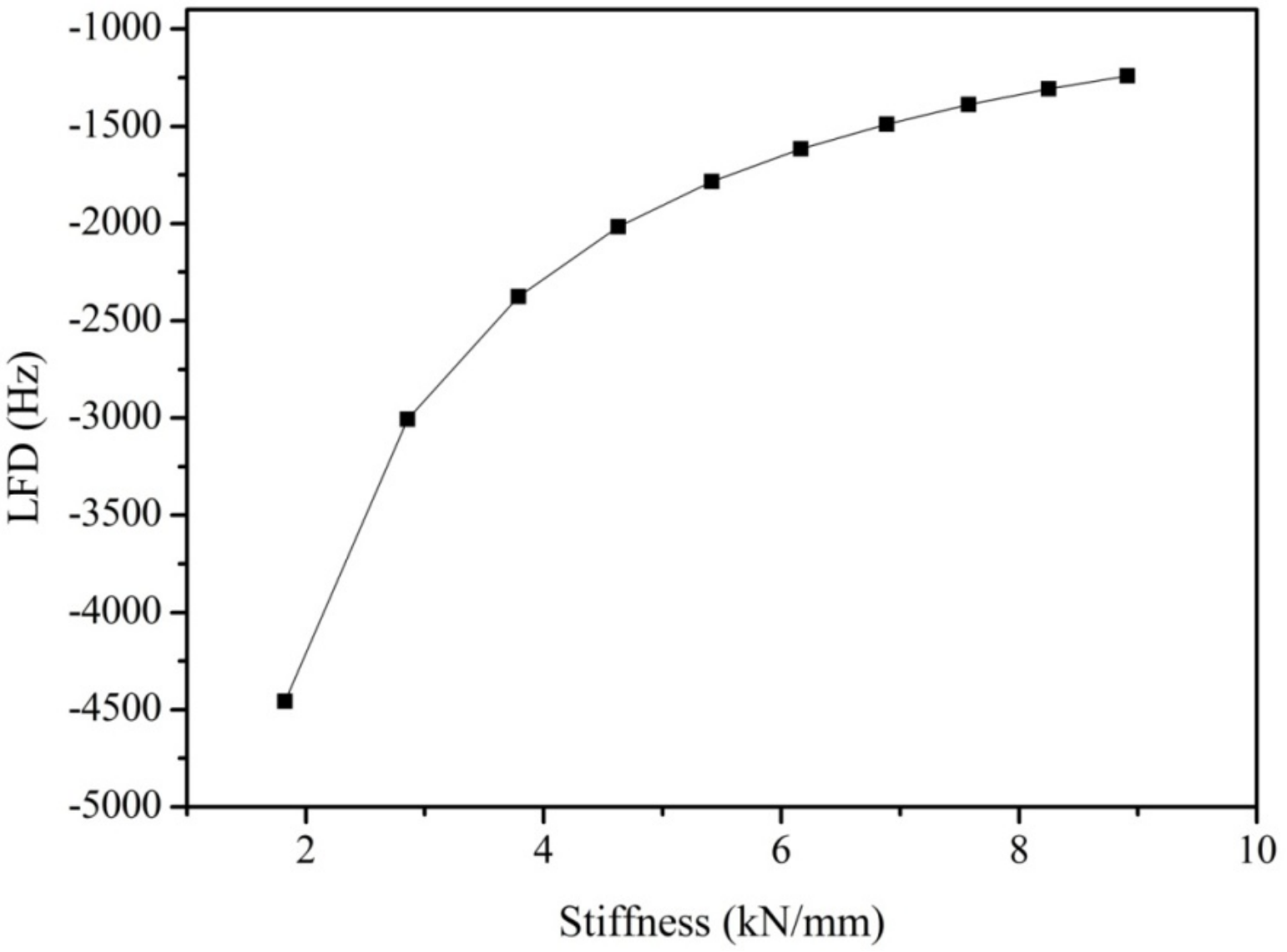

Fig. 19. LFD as a function of stiffness of the helium vessel. These calculation are for the case when there is no stiffener ring.

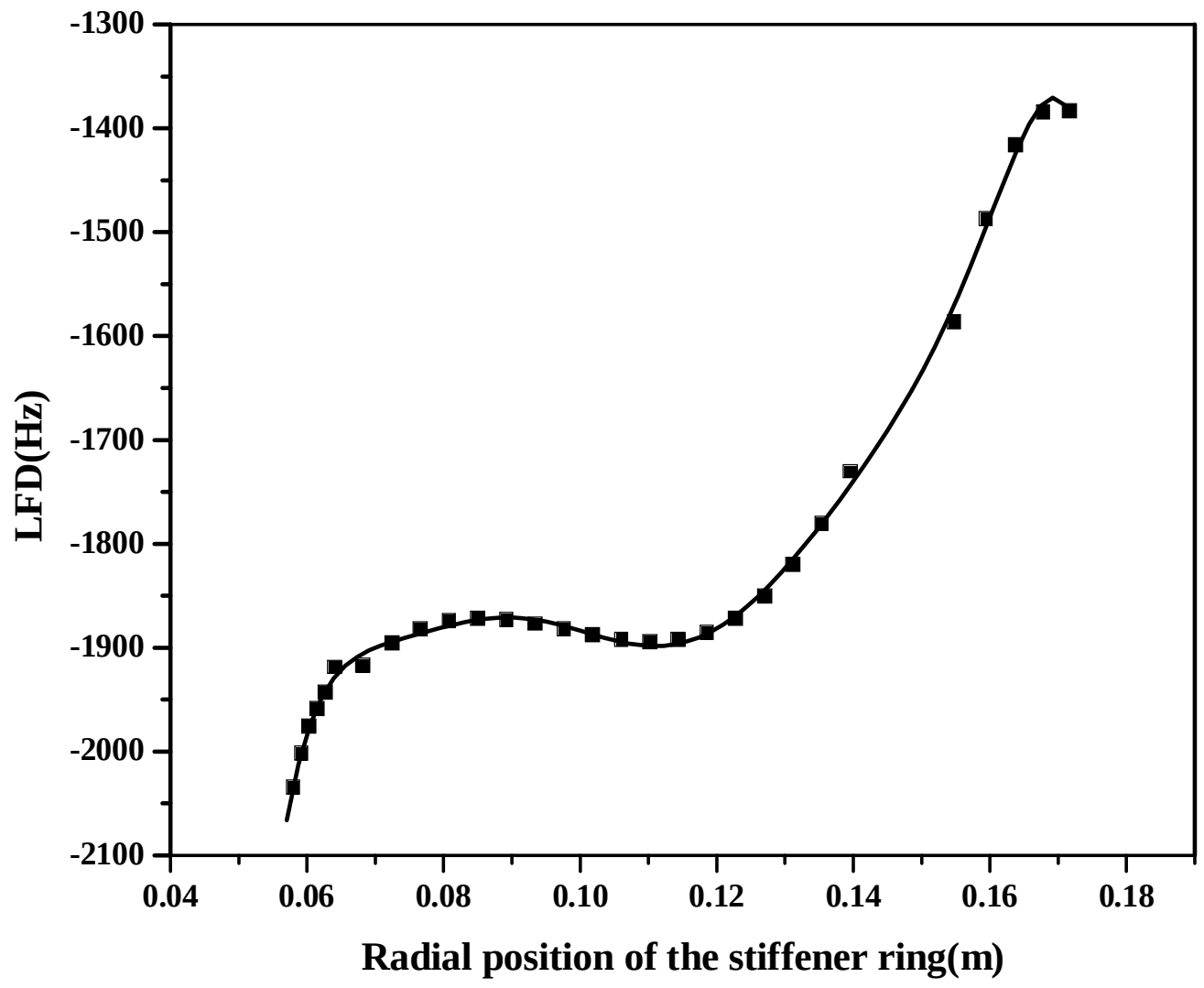

Fig. 20. LFD as a function of the radial position of the stiffener rings. Stiffness of the helium vessel is taken as 5 kN/mm in these calculations.

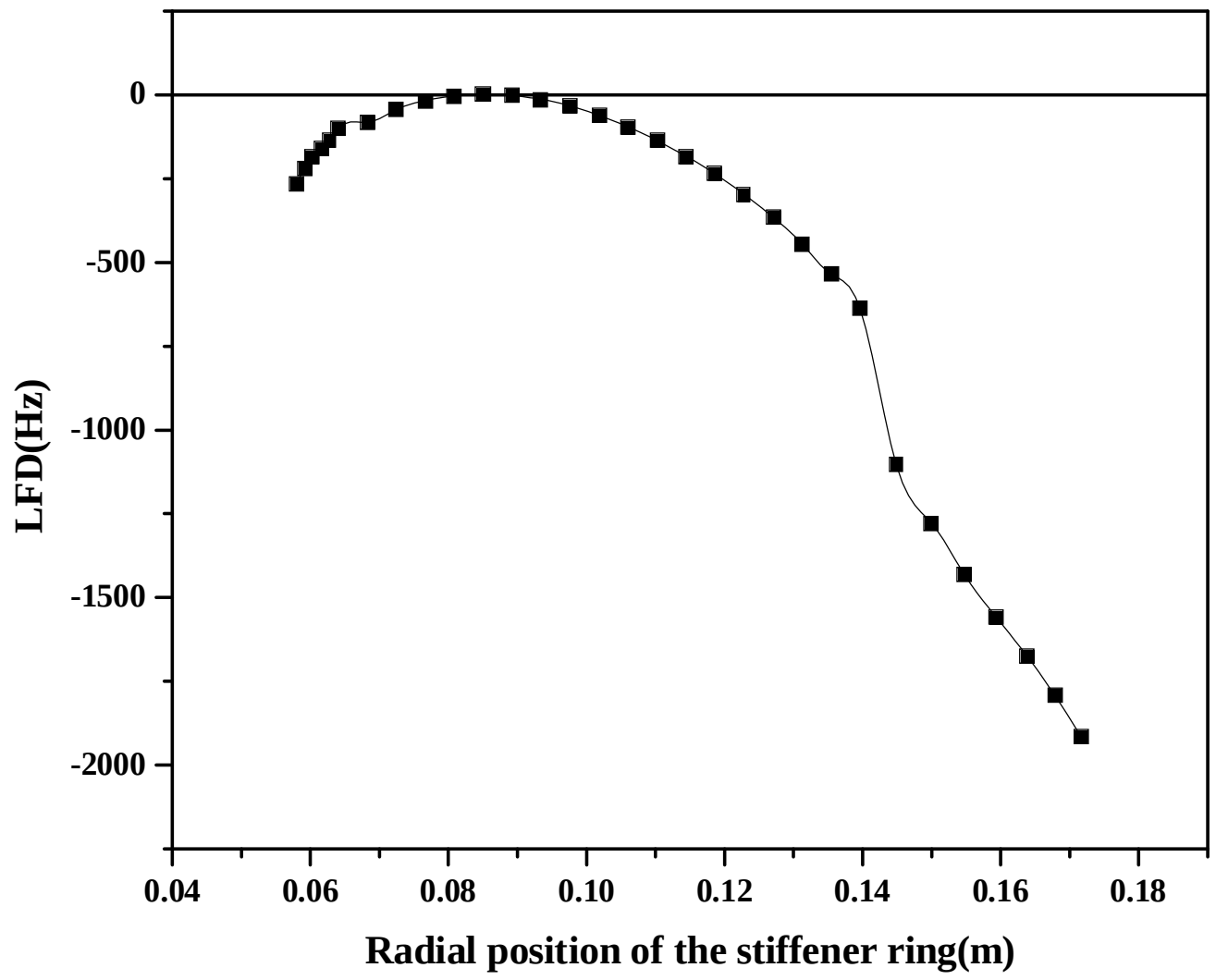

Fig. 21. LFD as a function of the radial position of the stiffener rings. Stiffness of the helium vessel is taken as 5 kN/mm. The compensation due to cavity elongation provided by a displacement of 8.3μm of the helium vessel is taken into account here.

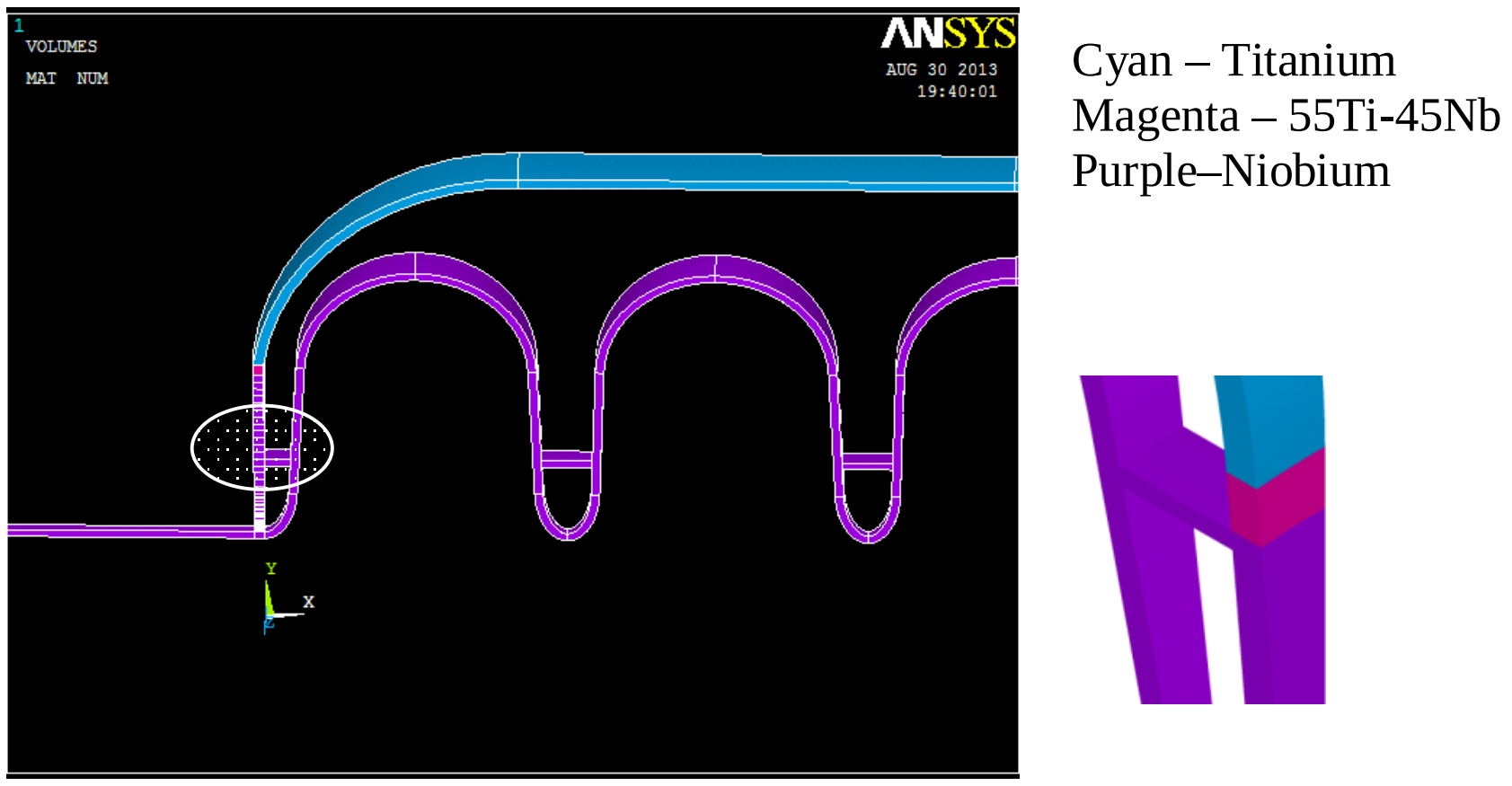

Fig. 22.Modified model of cavity with helium vessel.

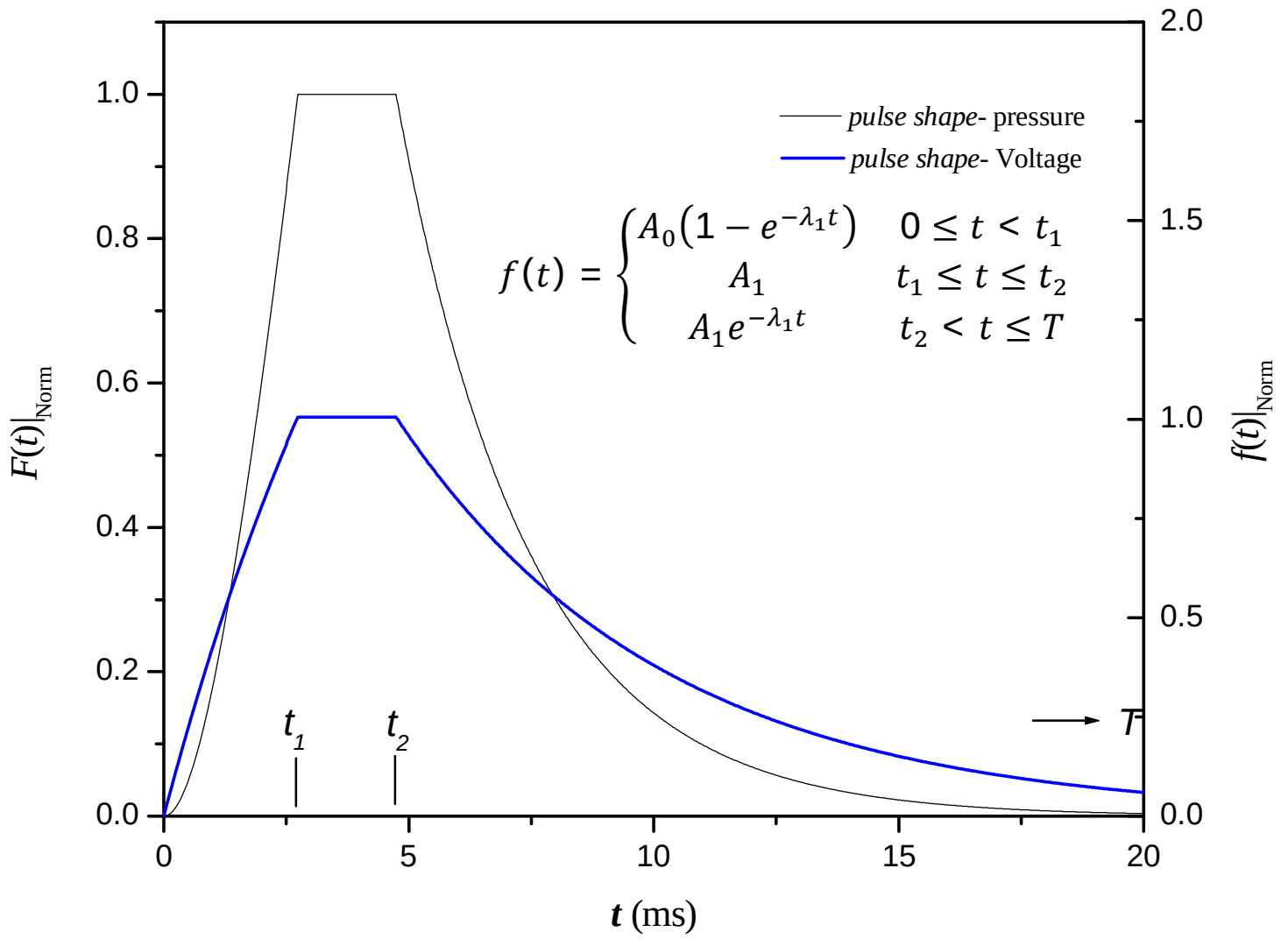

Fig. 23 .The normalized amplitude of the Lorentz Pressure Pulse (in black) and the input RF pulse (in blue) as a function of time is shown by the black line and blue line respectively.

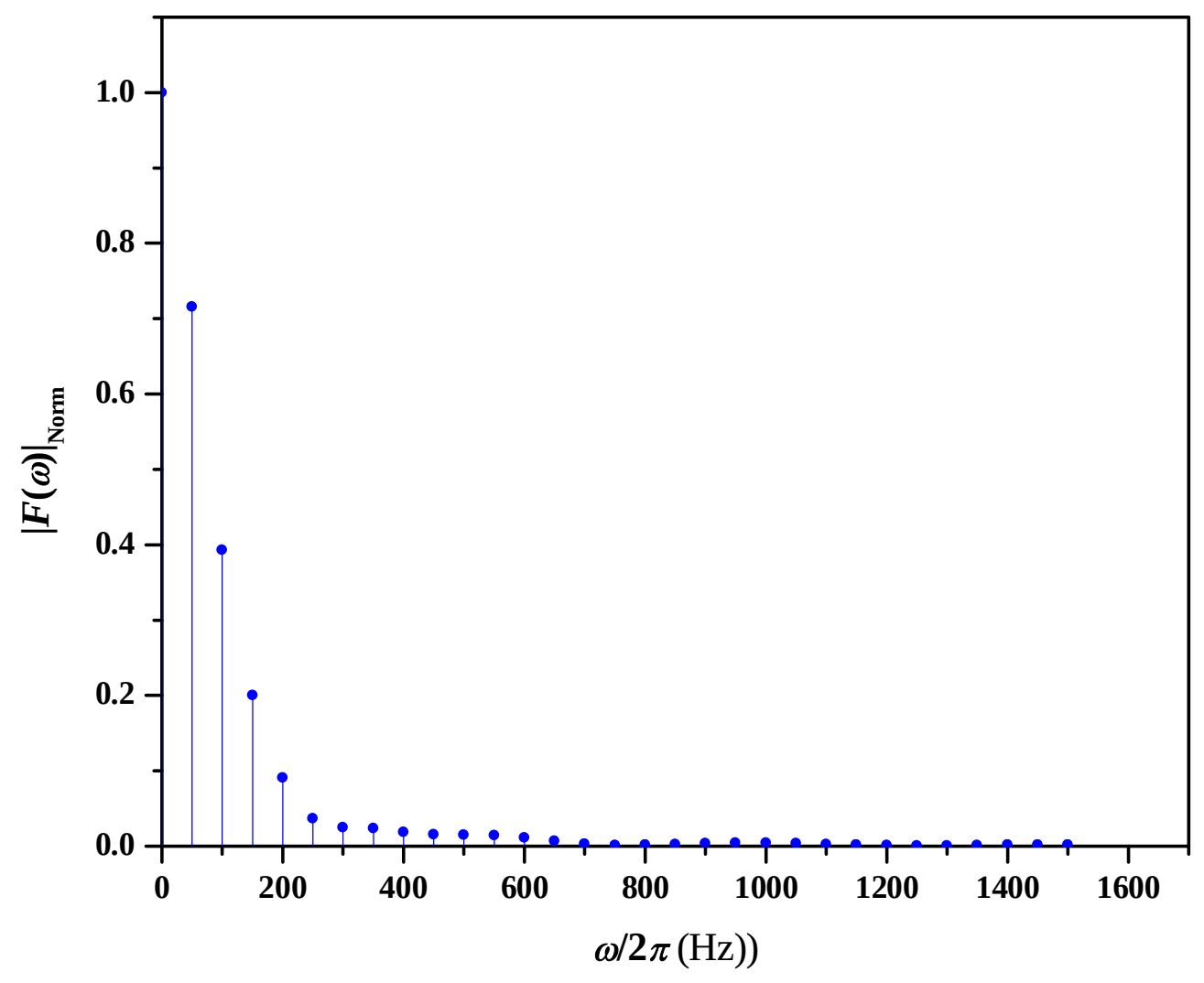

Fig. 24.Fourier spectrum of the normalized Lorentz Pressure Pulse Shape for calculation of dynamic LFD. Here $F(\omega)|_{Norm}$ is the normalized amplitudes correspond to the participating frequencies.

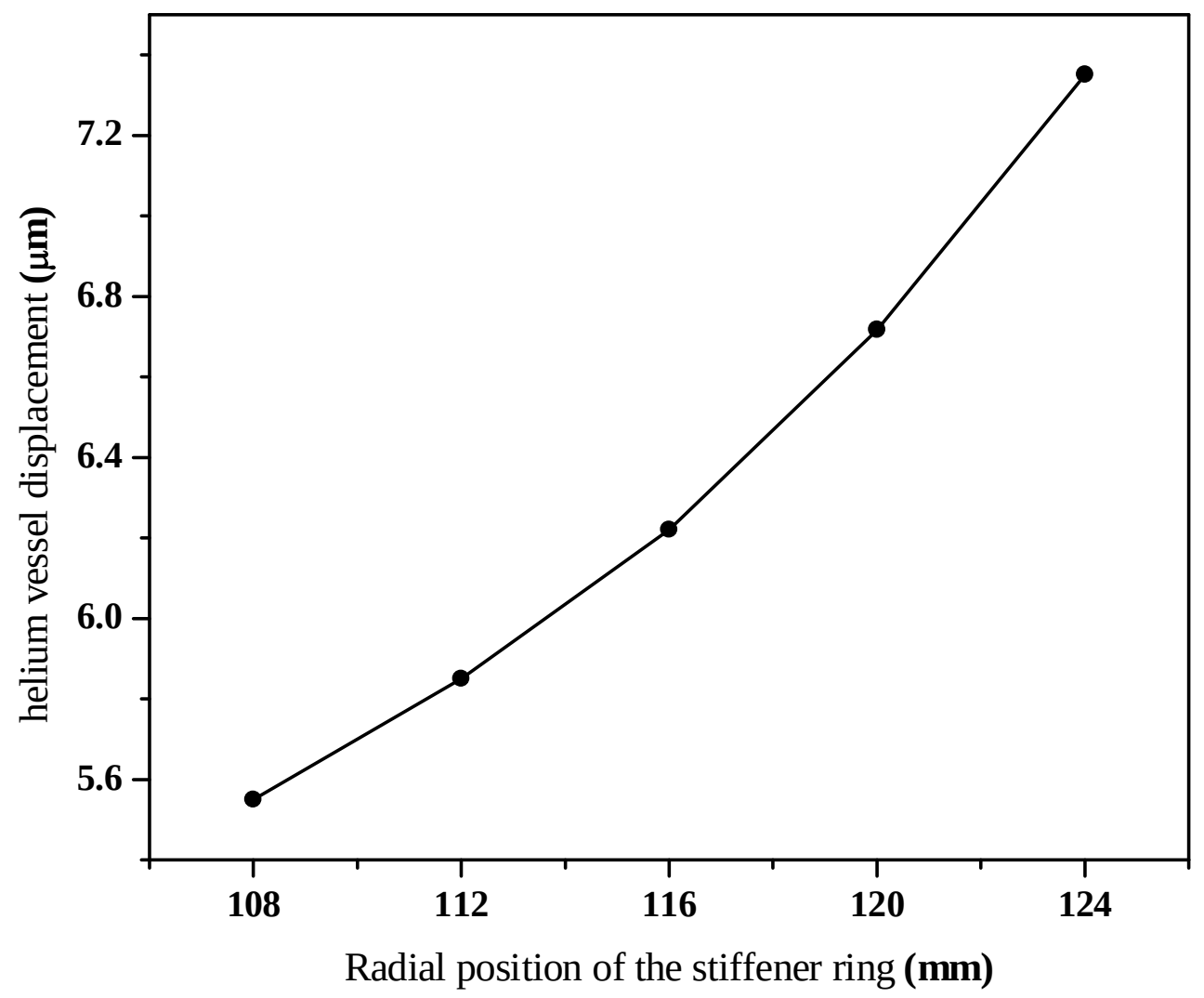

Fig. 25. Helium vessel displacement required to compensate for LFD,as a function of the radial position of the stiffener rings.

**TABLES:**

TABLE I
OPTIMIZED PARAMETERS FOR THE MID CELL GEOMETRY.

| Parameter | Magnitude | Unit |
|---|---|---|
| $R_{iris}$ | 44.00 | mm. |
| $R_{eq}$ | 195.591 | mm. |
| $L$ | 70.336 | mm. |
| $A$ | 52.64 | mm. |
| $B$ | 55.55 | mm. |
| $a$ | 15.280 | mm. |
| $b$ | 28.830 | mm. |
| $\alpha$ | $88^0$ | |

TABLE II
RF PARAMETERS OF THE OPTIMIZED MID CELL GEOMETRY FOR THE π MODE OPERATION.

| RF Parameter | Magnitude | Unit |
|---|---|---|
| Frequency | 650 | MHz |
| Transit-time factor(T) | 0.766 | |
| $Q_0$ | $> 1\times10^{10}$ | |
| $R/Q_0$ | 65.33 | Ω. |
| $E_{pk}$ | 36.05 | MV/m |
| $B_{pk}$ | 69.8 | mT |

TABLE III
OPTIMIZED PARAMETERS FOR THE CAVITY GEOMETRY.

| Parameter | Magnitude | | Unit |
|---|---|---|---|
| | End Cell-A | End Cell-B | |
| $R_{iris}$ | 44.00 | 44.00 | mm. |
| $R_{eq}$ | 195.591 | 195.591 | mm. |
| $L$ | 71.550 | 71.240 | mm. |
| $A$ | 52.640 | 52.250 | mm. |
| $B$ | 55.550 | 55.550 | mm. |
| $a$ | 15.280 | 15.280 | mm. |
| $b$ | 28.830 | 28.830 | mm. |

TABLE IV

RF PARAMETERS OF THE OPTIMIZED MID CELL GEOMETRY FOR THE π MODE OPERATION.

| RF Parameter | Magnitude | Unit |
|---|---|---|
| Frequency | 650 | MHz. |
| Transit-time factor(T) | 0.710 | |
| $Q_0$ | $> 1\times10^{10}$ | |
| $R/Q_0$ | 327.910 | Ω. |
| $E_{pk}$ | 36.66 | MV/m. |
| $B_{pk}$ | 70.83 | mT |

TABLE V

DETAILS OF THE PROMINENT DIPOLE MODES SUPPORTED BY THE CAVITY.

| Mode frequency $f$ (in MHz) | $\beta_p$ | $Q$ | $R_\perp/Q$ ($\Omega/m^2$) | $I_{th}$ (in mA) |
|---|---|---|---|---|
| 961.982 | 0.760 | $2.165\times10^9$ | 44313.698 | 1.007 |
| 965.852 | 0.659 | $2.229\times10^9$ | 28279.583 | 0.701 |
| 971.217 | 0.588 | $2.335\times10^9$ | 16987.028 | 0.876 |
| 976.495 | 0.532 | $2.465\times10^9$ | 8868.9049 | 1.335 |
| 980.104 | 0.494 | $2.574\times10^9$ | 4606.5459 | 2.191 |
| 1296.33 | 0.731 | $2.165\times10^9$ | 44313.698 | 2.319 |

TABLE VI

PARTICIPATING STRUCTURAL MODES OF THE 5-CELL CAVITY.

| $r_{stiffener}$(mm) | $f_1$(Hz) | $f_2$(Hz) | $f_3$(Hz) | $f_4$(Hz) | $f_5$(Hz) |
|---|---|---|---|---|---|
| 124.00 | 265.07 | 426.48 | 576.19 | 713.59 | 749.42 |
| 120.00 | 244.87 | 414.74 | 564.03 | 696.97 | 759.72 |
| 116.00 | 226.00 | 397.09 | 550.89 | 681.37 | 760.72 |
| 112.00 | 208.76 | 375.45 | 538.55 | 662.98 | 750.70 |
| 108.00 | 193.18 | 352.27 | 526.39 | 641.40 | 732.29 |